\documentstyle[12pt,epsfig,amssymb]{article}
\begin{document}
\setlength{\baselineskip}{0.75cm}
\setlength{\parskip}{0.45cm}
\begin{titlepage}
\begin{flushright}
DO-TH 97/20 \linebreak
September 1997 \linebreak
revised February 1998 
\end{flushright}
\vskip 0.8in
\begin{center}
{\Large\bf Dark matter constraints in the minimal \\ 
  and nonminimal SUSY standard model}

\vspace{1.2cm}
\large
A.\ Stephan\\
\vspace{0.5cm}
\normalsize
Universit\"{a}t Dortmund, Institut f\"{u}r Physik, \\
\vspace{0.1cm}
D-44221 Dortmund, Germany \\
\vspace{1.6cm}
{\bf Abstract} 
\vspace{-0.3cm}
\end{center}

\noindent
We determine the allowed parameter space and the particle spectra 
of the minimal SUSY standard model (MSSM) and 
nonminimal SUSY standard model (NMSSM) imposing correct 
electroweak gauge-symmetry breaking and recent experimental constraints. 
The parameters of the models are evolved with the SUSY RGEs
assuming universality at the GUT scale.
Applying the new unbounded from below (UFB) constraints we can exclude 
the LSP singlinos and light scalar and pseudoscalar Higgs singlets 
of the NMSSM.
This exclusion removes the experimental possibility to distinguish between 
the MSSM and NMSSM via the recently proposed search for an additional 
cascade produced in the decay of the bino into the LSP singlino. 
Furthermore the effects of the dark matter condition for the MSSM
and NMSSM are investigated and the differences concerning 
the parameter space, the SUSY particle and Higgs sector are discussed.
\end{titlepage}
%
%
\section{Introduction}

\noindent
If the standard model (SM) is embedded in a grand unified theory (GUT),
finetuning is needed in order to have a Higgs mass of the order of the 
electroweak scale and not the GUT scale. 
The quadratic divergences in the radiative corrections to the 
Higgs mass are the reason for this naturalness problem in the SM.
SUSY models have the advantage of being free of quadratic divergences 
due to cancellations in Feynman graphs with particles and their 
supersymmetric partners.
The simplest SUSY extension of the SM is the 
minimal SUSY standard model (MSSM) \cite{nilles, haber}
with two Higgs doublets $H_1$ and $H_2$.
However the MSSM has also a naturalness problem why the 
$\mu$-parameter of the term $\mu \: H_1 \: H_2$ in the superpotential 
should be much smaller than the Planck mass.
The nonminimal SUSY standard model (NMSSM) \cite{ellis}
is a minimal extension of the MSSM.
In this model a Higgs singlet N is added to 
the Higgs sector of the MSSM.
Moreover the $\mu$-parameter of the MSSM can be dynamically generated 
via $\mu = \lambda x$ with the term $\lambda \: H_1 \: H_2 \: N$ 
in the superpotential of the NMSSM, if the Higgs singlet acquires a 
vaccum expectation value (VEV) $\langle N \rangle = x$. 
SUSY models with Higgs singlets can be derived from superstring 
inspired $E_6$ or $SU(5) \times  U(1)$ GUT models
and they offer the possibility of spontaneous breaking 
of the CP symmetry.
The discrete ${\Bbb Z}_{\, 3}$ symmetry of the NMSSM causes a
domain wall problem.
In the early universe the ${\Bbb Z}_{\, 3}$ symmetry
is spontaneously broken during the electoweak phase transition.
Domains of different degenerate vacua develop which are 
separated by domain walls.
Since these domain walls would dominate the energy density of 
the universe and disturb primordial nucleosynthesis, 
they have to disappear before the onset of nucleosynthesis.
Possible solutions to the domain wall problem and the hierarchy 
problem reintroduce the terms $\mu \: H_1 \: H_2$ and $\mu' \: N^2$ 
in the superpotential of the NMSSM and possess a 
gauged R-symmetry or a target space duality symmetry 
at the Planck scale \cite{abelw}.
In these more general models the NMSSM is approximated for
very small values of the parameters $\mu$ and $\mu'$. 
\newline
\noindent
The addition of a Higgs singlet superfield leads to
two extra Higgses and one extra neutralino in the NMSSM
compared to the particle spectrum of the MSSM.
Since the Higgs singlets and the singlino $\tilde{N}$
can mix with the other Higgses and neutralinos,
the phenomenology of the NMSSM may be modified compared 
to the MSSM.
The phenomenology of the NMSSM has been investigated in
several papers \cite{ellwglmin} - \cite{franke}.
In SUSY models with R-parity conservation the lightest 
SUSY particle (LSP) cannot decay and is therefore stable.
Especially a neutral LSP like the lightest neutralino
is a good candidate for cold dark matter.
In the MSSM the dark matter neutralino is mostly a bino.
In the NMSSM there exists the alternative of a 
dark matter neutralino with a larger singlino portion.
The cosmology of these dark matter singlinos has been discussed 
in \cite{greene} at a given fixed low energy scale.
The experimentally and cosmologically allowed
parameter space of the NMSSM and the dark matter neutralinos
have been investigated in \cite{abeln} using SUSY RG evolutions.
\newline
\noindent
In the present paper we compare the allowed parameter space 
and the particle spectra of the MSSM and NMSSM.
Assuming universality at the GUT scale the parameters of the models 
are evolved to the electroweak scale with the SUSY RGEs
(given in \cite{savoy} and Appendix A2 for the NMSSM).
The Higgs potential is minimized and additional theoretical constraints
and recent experimental bounds are imposed.
Besides charge and color breaking (CCB) constraints we include here
new unbounded from below (UFB) constraints \cite{casas}.
Furthermore the consequences of the dark matter condition for 
the MSSM and NMSSM are investigated.
In the present paper we give a more detailed analysis of the two models 
and the dark matter condition than in our previous paper \cite{ich}. 
\newline
In section 2 we introduce the NMSSM. 
The additional theoretical and 
experimental constraints are discussed in section 3.
In section 4 we describe the dark matter constraints.
The procedure of finding solutions of the NMSSM is explained 
in section 5.
In section 6 the parameter space and the particle spectra 
of the MSSM and NMSSM are investigated without and with
the dark matter condition.
In Appendix A1 we give the complete formulas for the neutralino 
annihilation cross section in the NMSSM which are needed  
to calculate the relic neutralino density.
In Appendix A2 the SUSY RGEs \cite{savoy} for the NMSSM are listed.

\section{The NMSSM}

\noindent
The Higgs sector of the MSSM consists of two Higgs doublets
$H_1$ and $H_2$.
In the NMSSM \cite{ellis} an additional Higgs singlet $N$
is introduced.
The superpotential of the NMSSM contains only Yukawa terms with
dimensionless couplings
\begin{eqnarray}
   W = h_d \: H_1^T \: \epsilon \: \tilde{Q} \: \tilde{D} 
     - h_u \: H_2^T \: \epsilon \: \tilde{Q} \: \tilde{U} 
     + \lambda \: H_1^T \: \epsilon \: H_2 \: N 
     - \frac{1}{3} \: k \: N^3.  
\end{eqnarray}
\noindent
$\tilde{Q}$ is a squark doublet. $\tilde{U}$ and $\tilde{D}$ are up-type 
and down-type squark singlets.
$\epsilon$ is the antisymmetric tensor with $\epsilon_{1 2} \: = \: 1$.
The SUSY breaking is parametrized by different types of SUSY 
soft breaking terms in the lagrangian:
\begin{eqnarray}
{\cal L}_{soft} & = &  - \, m^2_{H_1} \: |H_1|^2 - m^2_{H_2} \: |H_2|^2
   - m^2_N \: |N|^2 - m^2_Q \: |\tilde{Q}|^2
   - m^2_U \: |\tilde{U}|^2 - m^2_D \: |\tilde{D}|^2  \nonumber \\
   &  &  - \, h_d \: A_d \: H_1^T \: \epsilon \: \tilde{Q} \: \tilde{D} 
   + h_u \: A_u \: H_2^T \: \epsilon \: \tilde{Q} \: \tilde{U} 
   \nonumber \\
   &  &  + \, \lambda \: A_{\lambda} \: H_1^T \: \epsilon \: H_2 \: N 
   + \frac{1}{3} \: k \: A_k \: N^3  \nonumber \\ 
   &  &  + \, \frac{1}{2} \: M_3 \: \lambda_3 \: \lambda_3
   + \frac{1}{2} \: M_2 \: \lambda^a_2 \: \lambda^a_2
   + \frac{1}{2} \: M_1 \: \lambda_1 \: \lambda_1 + h.c. 
\end{eqnarray}
\noindent
In SUSY models the tree-level scalar potential is given by
the formula:
\begin{eqnarray}
   & V_{tree} = \frac{1}{2} \sum (D^a)^2 + \sum F^{\ast}_i \, F_i 
   + V_{soft} &  \label{spot}  \\
   & D^a = g_a \, A^{\ast}_i \, T^a_{i j} \, A_j, & \label{da}  \\ 
   & F_i = - \left( \frac{\displaystyle \partial W}
   {\displaystyle \partial A_i} \right)^{\ast} & \label{fi} 
\end{eqnarray}
\noindent
where $A_i$ are scalar fields and $T^a_{i j}$ are the 
gauge group generators.
The part of the scalar potential with only Higgs fields 
corresponds to the Higgs potential.
The neutral part of the tree-level Higgs potential is given by:
\begin{eqnarray}
 V_{tree}^{Higgs} & = & m^2_{H_1} \: |H^0_1|^2 + m^2_{H_2} \: |H^0_2|^2 
   + m^2_N \: |N|^2 + \lambda^2 \: (|H^0_1|^2 + |H^0_2|^2) \: |N|^2
   + \lambda^2 \: |H^0_1|^2 \: |H^0_2|^2  \nonumber \\
   &  &  + \: k^2 \: |N|^4 - \lambda \: k \: H^0_1 \: H^0_2 \: N^{\ast 2}
   - \lambda \: A_{\lambda} \: H_1^0 \: H_2^0 \: N 
   - \frac{1}{3} \: k \: A_k \: N^3  \nonumber \\
   & &  + \: \frac{1}{8} \: (g_y^2 + g_2^2) \: (|H^0_1|^2 - |H^0_2|^2)^2
   + h.c.  \label{vhiggs}
\end{eqnarray}
\noindent
In order to calculate the Higgs VEVs reliably we use the effective 
1 loop Higgs potential consisting of the tree-level Higgs potential 
and the radiative corrections
\begin{eqnarray}
 V_{1 loop}^{Higgs} = V_{tree}^{Higgs} + V_{rad}^{Higgs}. 
 \label{vh1l}
\end{eqnarray}
\noindent
The radiative corrections \cite{ellwp} to the effective Higgs potential 
arising from all particles and antiparticles with field-dependent mass 
$m_i$, spin $S_i$ and color degrees of freedom $C_i$ are given by
\begin{eqnarray}
   V_{rad}^{Higgs} = \frac{1}{64 \: \pi^2} \sum_{i}
   \: C_i \: (-1)^{2 S_i} \: (2 S_i + 1) \: m_i^4 \:
   \: \ln(\frac{m_i^2}{Q^2}). 
   \label{vrad}
\end{eqnarray}
\noindent
$Q$ is the renormalisation scale which we take to be equal to 
the electroweak scale $M_{weak} \simeq  100 \: GeV$.
Here we only consider the important contributions of the top quark and 
stops $\tilde{t}_{1,2}$.
The electroweak gauge-symmetry $SU(2)_I \times U(1)_Y$
is spontaneously broken to the electromagnetic 
gauge-symmetry $U(1)_{em}$ by the Higgs VEVs
$\langle H_i^0 \rangle = v_i$  with $i = 1,2$ 
and $\langle N \rangle = x$.
The local minimum of the effective 1 loop Higgs potential
is determined by the three minimum conditions for the VEVs:
\begin{eqnarray}
   \frac{1}{2} \: m_Z^2 = 
   \frac{m_{H_1}^2 + \Sigma^1 
   - (m_{H_2}^2 + \Sigma^2) \: \tan^2 \beta }  
   {\tan^2 \beta - 1}
   - \lambda^2 x^2  \label{nmin1}
\end{eqnarray}
\begin{eqnarray}
   \sin(2 \beta) = 
   \frac{2 \: \lambda \: x \: (A_{\lambda} + k \: x)}
   {m_{H_1}^2 + \Sigma^1 + m_{H_2}^2 + \Sigma^2 
    + 2 \: \lambda^2 \: x^2 + \lambda^2 \: v^2}  \label{nmin2}
\end{eqnarray}
\begin{eqnarray}
   (m_N^2 + \Sigma^3) \: x^2 - k \: A_k \: x^3 + 2 \: k^2 \: x^4
   + \lambda^2 \: x^2 \: v^2
   - \frac{1}{2} \: (A_{\lambda} + 2 \: k \: x) \:
   \lambda \: x \: v^2 \: \sin(2 \beta) \: = \: 0  \label{nmin3}
\end{eqnarray}
\noindent
where $v = \sqrt{v_1^2 + v_2^2} = 174 \: GeV$,  
$\tan \beta = v_2/v_1$ and 
$\Sigma^i = \partial V^{Higgs}_{rad}/\partial v_i^2$.
\newline
\noindent
The neutralinos are Majorana particles and mixed states of the
gauginos and Higgsinos.
The mass and composition of the neutralinos is defined by 
the following part of the lagrangian
\begin{eqnarray}
   {\cal L} = -\frac{1}{2} \Psi^T M \Psi + h.c.  
\end{eqnarray}
\begin{eqnarray}
   \Psi^T = (-i \lambda_1,-i \lambda^3_2,
   \Psi^0_{H_1},\Psi^0_{H_2},\Psi_N). \label{nbas} 
\end{eqnarray}
\noindent
In this basis the symmetric mass matrix $M$ of 
the neutralinos has the form:  
\[ \left( \begin{array}{ccccc}

   M_1  &  0  &  - m_Z \: \sin \theta_W \: \cos \beta  &
   m_Z \: \sin \theta_W \: \sin \beta  &  0  \\

   0  &  M_2  &  m_Z \: \cos \theta_W \: \cos \beta  &
   - m_Z \: \cos \theta_W \: \sin \beta  &  0  \\

   - m_Z \: \sin \theta_W \: \cos \beta  &  
   m_Z \: \cos \theta_W \: \cos \beta  &  0  &
   \lambda \: x  &  \lambda \: v_2  \\

   m_Z \: \sin \theta_W \: \sin \beta  & 
   - m_Z \: \cos \theta_W \: \sin \beta  &  
   \lambda \: x  &  0  &  \lambda \: v_1  \\
 
   0  &  0  &  \lambda \: v_2  &  \lambda \: v_1  &  - 2 \: k \: x  

   \end{array}\right). \]
\noindent
With $\mu = \lambda \: x$ the first $4 \times 4$ submatrix 
recovers the mass matrix of the MSSM \cite{haber}.  
The mass of the neutralinos is here obtained by diagonalising
the mass matrix $M$ with the orthogonal matrix $N$.
(Then some mass eigenvalues may be negative \cite{bartln}.)
\begin{eqnarray}
   {\cal L} = -\frac{1}{2} \: m_i \:
   \overline{\tilde{\chi}^0_i} \: \tilde{\chi}^0_i 
\end{eqnarray}
\begin{eqnarray}
   \tilde{\chi}^0_i = \left(\begin{array}{cc}
   \chi^0_i  \\ \overline{\chi}^0_i 
   \end{array}\right) \:\: \mbox{with} \:\:
   \chi^0_i = N_{ij} \Psi_j \:\:\: and \:\:\:
   M_{diag} = N \: M \: N^{T}.  \label{xi}
\end{eqnarray}
\noindent
The neutralinos $\tilde{\chi}^0_i \:\:\: (i = 1-5)$ are ordered
with increasing mass $|m_i|$, thus $\tilde{\chi}^0_1$ is the 
LSP neutralino.
The matrix elements $N_{i j} \:\:\: (i,j = 1-5)$ describe the 
composition of the neutralino $\tilde{\chi}^0_i$ in the 
basis $\Psi_j$ (Eq.(\ref{nbas})).
For example the bino portion of the LSP neutralino is given
by $N^2_{1 1}$ and the singlino portion of the LSP neutralino
by $N^2_{1 5}$.
The diagonalising of the mass matrix $M$ leads to a 
polynomial equation of degree 5, which cannot be solved 
analytically in general.
Therefore we calculate the eigenvalues and eigenvectors
of the symmetric mass matrix $M$ numerically 
by using a subroutine in the NAG Fortran Library.
\newline
\noindent
The possibility of gauge coupling unification is an important
motivation for SUSY GUTs.
In contrast to non-supersymmetric GUTs \cite{ellisgut},
the gauge couplings in SUSY GUTs are allowed to unify 
$g_a(M_X) = g_5 \simeq 0.72$ at a scale 
$M_X \simeq 1.6 \times 10^{16} \: GeV$.
Within a minimal $N = 1$ supergravity framework universality 
naturally arises for the SUSY soft breaking parameters:
\begin{eqnarray}
   \begin{array}{l}
 m_i(M_X) = m_0  \\
 M_a(M_X) = m_{1/2}  \\
 A_i(M_X) = A_0  \hspace*{1.5cm} (A_{\lambda}(M_X) = - A_0).   
   \end{array}  \label{univ}
\end{eqnarray}
\noindent
The Yukawa couplings at the GUT scale take the values 
$\lambda(M_X) = \lambda_0$, $k(M_X) = k_0$ and $h_{t}(M_X) = h_{t 0}$.
With the SUSY RGEs (\cite{savoy}, Appendix A2) the SUSY soft breaking 
parameters and the Yukawa couplings are evolved from the GUT scale 
to the electroweak scale $M_{weak}$. 
At $M_{weak}$ we minimize the effective 1 loop Higgs potential
$V_{1 loop}^{Higgs}$ (Eq.(\ref{vh1l})).

\section{Theoretical and experimental constraints}

\noindent
The solutions of the SUSY RGEs (\cite{savoy}, Appendix A2)
have to fulfill several theoretical and experimental constraints.
The minimum of the effective 1 loop Higgs potential
$V_{1 loop}^{Higgs}$ (Eq.(\ref{vh1l}))
should be not only a local minimum but also a stable minimum.
A local minimum is stable against tunnelling into a
lower minimum, if its lifetime is large enough
(of the order of the age of the universe).  
A global minimum is certainly stable, but it may be difficult
to prove, that it is really a global minimum.
Here we only investigate, whether the physical minimum
is lower than the unphysical minima of 
$V_{1 loop}^{Higgs}(v_1, v_2, x)$ with at least one 
vanishing VEV \cite{ellwglmin}.
\newline
\noindent
The conservation of charge and color would be broken by 
slepton and squark VEVs.
To avoid a scalar potential with such a minimum, the trilinear 
couplings have to fulfill the following conditions \cite{savoy}:
\begin{eqnarray}
  A_{u_i}^2 \leq 3 (m_{H_2}^2 + m_{Q_i}^2 + m_{U_i}^2 ) \: \:
  \mbox{at scale} \: \: Q \sim A_{u_i}/h_{u_i}  \label{aui} \\    
  A_{d_i}^2 \leq 3 (m_{H_1}^2 + m_{Q_i}^2 + m_{D_i}^2 ) \: \:      
  \mbox{at scale} \: \: Q \sim A_{d_i}/h_{d_i}  \label{adi} \\
  A_{e_i}^2 \leq 3 (m_{H_1}^2 + m_{L_i}^2 + m_{E_i}^2 ) \: \:     
  \mbox{at scale} \: \: Q \sim A_{e_i}/h_{e_i}.  \label{aei} 
\end{eqnarray}
\noindent
In \cite{casas} improved charge and color breaking (CCB) constraints
and unbounded from below (UFB) constraints are analysed
for the MSSM.
The by far strongest of these new constraints is the UFB-3 constraint
(Eq.(33) in the first publication of Ref.\cite{casas}),
in contrast to the remaining ones.
To apply the UFB-3 constraint to the NMSSM we have to modify 
the formulas in \cite{casas}:
In the NMSSM a dangerous field direction of the tree-level 
scalar potential $V_{tree}$ (Eq.(\ref{spot})) occurs 
for VEVs of the neutral Higgs $H_2^0$, 
the electron sneutrino $\tilde{\nu}_e$ denoted by $L_1 = \tilde{L}_1^0$ 
and the staus $\tilde{\tau}^-_L$ and $\tilde{\tau}^+_R$ denoted by
$\tilde{L}_3^-$ and $\tilde{E}_3$.
The VEVs of the staus are introduced in order to cancel the 
F-term of $H_1^0$:
\begin{eqnarray}
  \left| \: \frac{\partial \: W}{\partial \: H_1^0} \: \right|^2 =
  |\: \lambda \: x \: H^0_2 + h_{\tau} \: \tilde{L}_3^- 
   \: \tilde{E}_3 \: |^2 = 0. \label{fter}
\end{eqnarray}
\noindent
For the VEVs of the staus we choose the relation 
$e = \tilde{L}_3^- = \tilde{E}_3$.
In analogy to the MSSM we have replaced in Eq.(\ref{fter}) $\mu$ by
$\lambda \: x$ where the Higgs singlet VEV $\langle N \rangle = x$
is determined by the physical minimum of the NMSSM Higgs potential. 
Since we consider here the scalar potential and not only 
the Higgs potential, the Higgs singlet VEV $x$ may be optimized further.
But here we don't investigate this more general case.  
The NMSSM tree-level scalar potential along the 
described field direction is given by:
\begin{eqnarray}
  & V_{tree} & = \: m^2_{H_2} \: |H_2^0|^2 
  + (m^2_{L_3} + m^2_{E_3}) \: |e|^2 + m^2_{L_1} \: |L_1|^2  \nonumber \\
  & &  + \: \frac{1}{8} \: (g_y^2 + g_2^2) \: 
  \left( |H^0_2|^2 + |e|^2 - |L_1|^2 \right)^2   \nonumber \\
  & &  + \: m^2_N \: x^2 - \frac{2}{3} \: k \: A_k \: x^3 + k^2 \: x^4. 
  \label{dpot}
\end{eqnarray}
\noindent
Now we can determine the dangerous field direction of 
$V_{tree}$ (Eq.(\ref{dpot})), which is given by:
\begin{eqnarray}
  |L_1|^2 = \frac{- 4 \: m^2_{L_1}}{g_y^2 + g_2^2} + |H_2^0|^2 + |e|^2,
  \label{dil1}
\end{eqnarray}
\begin{eqnarray}
  |e| = \sqrt{ \frac{|\lambda \: x|}{h_{\tau}} \: |H_2^0| }  \label{die}
\end{eqnarray}
\noindent
provided that 
\begin{eqnarray}
  |H_2^0| > \sqrt{ \frac{(\lambda \: x)^2}{4 \: h_{\tau}^2} 
          + \frac{4 \: m^2_{L_1}}{g_y^2 + g_2^2} } 
          \: - \: \frac{\lambda \: x}{2 \: h_{\tau}}. 
\end{eqnarray}
\noindent
For smaller values of $|H^0_2|$ we obtain $|L_1| = 0$.
In the field direction of Eq.(\ref{dil1}) and (\ref{die}) 
the tree-level scalar potential of the NMSSM reads:
\begin{eqnarray}
  V_{UFB-3} & = & (m^2_{H_2} + m^2_{L_1}) \: |H_2^0|^2 
  + \frac{|\lambda \: x|}{h_{\tau}} \: 
  (m^2_{L_3} + m^2_{E_3} + m^2_{L_1}) \: |H_2^0|  \nonumber \\
  & & - \: \frac{2 \: m^4_{L_1}}{g_y^2 + g_2^2}  
  + m^2_N \: x^2 - \frac{2}{3} \: k \: A_k \: x^3 + k^2 \: x^4. 
\end{eqnarray}
\noindent
For $|L_1| = 0$ the tree-level scalar potential is given by: 
\begin{eqnarray}
  & V_{UFB-3} & = \: m^2_{H_2} \: |H_2^0|^2 
  + \frac{|\lambda \: x|}{h_{\tau}} \: (m^2_{L_3} + m^2_{E_3}) \: |H_2^0|
  \nonumber \\
  & &  + \: \frac{1}{8} \: (g_y^2 + g_2^2) \: 
  \left( |H^0_2|^2 + \frac{|\lambda \: x|}{h_{\tau}} \: |H^0_2| \right)^2
  + m^2_N \: x^2 - \frac{2}{3} \: k \: A_k \: x^3 + k^2 \: x^4. 
\end{eqnarray}
\noindent 
Since the tree-level scalar potential $V_{UFB-3}$ should be larger 
than the physical minimum of the tree-level Higgs potential 
$V_{tree}^{Higgs}$ (Eq.(\ref{vhiggs})),
the UFB-3 constraint can be written as
\begin{eqnarray}
  V_{UFB-3}(Q = \hat{Q}) \: > \: (\: V_{tree}^{Higgs}(Q = M_S) \:)_{Min}.
  \label{ufb3}
\end{eqnarray}
\noindent
The tree-level Higgs potential $V_{tree}^{Higgs}$ 
(Eq.(\ref{vhiggs})) is minimized at the scale $Q = M_S$,
where the radiative corrections $V_{rad}^{Higgs}$ (Eq.(\ref{vrad})) 
are very small.
$M_S$ turns out to be a certain average of SUSY masses.
For the same reason the tree-level scalar potential $V_{UFB-3}$ 
is evaluated at the scale $Q = \hat{Q} \sim Max(g_2 \: |e|, 
g_2 \: |H_2^0|, h_t \: |H_2^0|, g_2 \: |L_1|, M_S)$.
Since $V_{UFB-3}(Q = \hat{Q})$ depends on $|H_2^0|$ the relation 
(\ref{ufb3}) should be fulfilled for all $|H_2^0|$ values below $M_X$. 

\noindent
Because we assume $h_t \gg h_b, h_{\tau}$,
we restrict $|\tan \beta \: |$ to be smaller than 20 \cite{ellwtnb}.
For the Higgs singlet VEV $x$ we choose the range
$|x| < 80 \: TeV$ \cite{ellwtnb, ellwn, king}. 
In the present paper we choose the parameters $m_0$ and $|m_{1/2}|$ 
to be smaller than $1500 \: GeV$.
This restriction is motivated by requiring the absence of too much
fine-tuning \cite{king},
which disfavors solutions with enormous sensitivity of $m_Z$ to 
the GUT scale parameters $h_{t0}$ or $\lambda_0$ and $k_0$. 
The dark matter constraints further restrict the allowed parameter region 
of $m_0$ and $|m_{1/2}|$ as shown in Fig.\ref{mmg}.   
\newline
\noindent
For the top quark pole mass we take the range 
$169 \: GeV < m_t < 181 \: GeV$ corresponding to the 
recent world average measurement \cite{top}. 
Since charged or colored massive stable particles are 
cosmologically disfavored LSPs, we assume the lightest
neutralino to be the LSP and impose the constraints
$m_{\tilde{\chi}^0_1} \: < \: m_{\tilde{\tau}_1}$ and 
$m_{\tilde{\chi}^0_1} \: < \: m_{\tilde{t}_1}$.
The SUSY particles in the NMSSM have to fulfill the following
experimental conditions:
$ m_{\tilde{\nu}} \geq 41.8 \: GeV \:\: \cite{PD} $,
$ m_{\tilde{e}_R} \geq 70 \: GeV \:\: \cite{sfer} $,
$ m_{\tilde{q}} \geq 176 \: GeV 
   \:\: ( \mbox{if} \: \: m_{\tilde{g}} < 300 \: GeV ) 
   \:\: \cite{PD} $ and
$ m_{\tilde{q}} \geq 70 \: GeV \:\: ( \mbox{for all} 
  \: \: m_{\tilde{g}} ) \:\: \cite{bisset} $,
\newline
$ m_{\tilde{t}_1} \geq 63 \: GeV \:\: \cite{sfer} $,
$ m_{\tilde{g}} \geq 173 \: GeV \:\: \cite{FERMILAB} $,
$ m_{\tilde{\chi}^+_1} \geq 85.4 \: GeV \:\: \cite{charg} $,
$ m_{H^+} \geq 52 \: GeV \:\: \cite{chiggs} $,
\newline
$m_{S_1} \geq 74 \: GeV$ \cite{higgs} or the coupling 
of the lightest scalar Higgs $S_1$ to the Z boson 
is reduced compared to the SM \cite{L3}
assuming visible decays with branching ratios
like the SM Higgs \cite{ellwn}, and
\[ \sum_{i,j} \Gamma(Z \rightarrow 
   \tilde{\chi}^0_i \tilde{\chi}^0_j) < 30 \: MeV \:\:
   \cite{diaz} \]
\[ \Gamma(Z \rightarrow 
   \tilde{\chi}^0_1 \tilde{\chi}^0_1) < 7 \: MeV \:\:
   \cite{diaz} \]
\[ BR(Z \rightarrow 
   \tilde{\chi}^0_i \tilde{\chi}^0_j) < 10^{-5} 
   \:\:,\: (i,j) \not= (1,1) \:\: \cite{diaz}. \]

\section{Dark matter constraints}

\noindent
Inflationary cosmological models are proposed to solve the 
flatness, horizon and magnetic monopole problem.
In these models the universe is flat with a density
equal to the critical density 
$\rho_{crit} = 1.054 \times 10^{-5} \: h^2_0 \: GeV / cm^3$
implying a cosmological density parameter of 
$\Omega = \rho / \rho_{crit} = 1$. 
The Hubble constant $H_0 = 100 \: h_0 \: km \: s^{-1} \: Mpc^{-1}$
is restricted by observations to the range $0.4 < h_0 < 1$
of the scaled Hubble constant $h_0$.
To be consistent with the observed abundances of the isotopes
$D$, $^3He$, $^4He$ and $^7Li$ the big bang nucleosynthesis model
restricts the baryon density to $\Omega_{baryons} \leq 0.1$.
The discrepancy between the cosmological density 
$\Omega = \rho / \rho_{crit} = 1$ and the baryon density
requires a huge amount of dark matter in the universe.
One of the favored theories to explain the structure formation
of the universe is the cold + hot dark matter model (CHDM)
\cite{davis}.
In this model the dark matter consists of 
hot dark matter (massiv neutrinos)
and cold dark matter (neutralinos) with
$\Omega_{hot} \simeq 0.3$ and $\Omega_{cold} \simeq 0.65$.
With the allowed range of the scaled Hubble constant $h_0$
the condition for the relic density of the neutralinos 
in the CHDM model reads
\begin{eqnarray}
   0.1 \leq \Omega_{\chi} h_0^2 \leq 0.65. 
\end{eqnarray}
\noindent
The relic neutralino density can be calculated by solving the 
Boltzmann equation for the early universe.
The Boltzmann equation describes the particle densities in the 
cosmic plasma governed by the annihilation cross section.
The annihilation cross section of the neutralinos can be expanded
in powers of the relative velocity $v$ of the 
non-relativistic neutralinos:
$\sigma_{ann} v = a + b \: v^2$.
The thermally averaged annihilation cross section 
\cite{drees} - \cite{kolb} is then given by 
$< \: \sigma_{ann} v \: > \: = \: x^{1.5} \: / \: (2 \: \sqrt{\pi}) 
\: \int_{0}^{\infty} \: dv \: v^2 \: \exp \{- \: x \: v^2 \: / \: 4 \} 
\: \sigma_{ann} v \: = a + 6 \: b \: / \: x$ 
with $x = |m_{\tilde{\chi}^0_1}| \: / \: T$.
During the expansion of the universe its temperature falls and
reaches the freeze-out temperature of the neutralinos $T_{fr}$,
at which the neutralinos drop out of the thermal equilibrium 
with the cosmic plasma. 
The freeze-out temperature of the neutralinos $T_{fr}$
follows from the equation \cite{drees} - \cite{griest}
\begin{eqnarray}
   x_{fr} = \ln \left(0.0764 \: |m_{\tilde{\chi}^0_1}| \: M_P \:
   \frac{< \: \sigma_{ann} v \: >}{\sqrt{g_{\ast} \: x_{fr}}} 
   \:c\:(2 + c)\right),  
\end{eqnarray}
\noindent
where $M_P = 1.221 \times 10^{19} \: GeV$ is the Planck mass,
$g_{\ast} \approx 81$ is the effective number of degrees
of freedom at $T_{fr}$ and $c = 1/2$.
With $J(x_{fr}) \: = \: \int_{x_{fr}}^{\infty} \: dx \: / \: x^2
\: < \: \sigma_{ann} v \: > \: = \: a \: / \: x_{fr}
+ 3 \: b \: / \: x_{fr}^2$ \cite{drees} - \cite{kolb}
the relic neutralino density 
\cite{drees} - \cite{griest} is given by 
\begin{eqnarray}
   \Omega_{\chi} h_0^2 = \frac{1.07 \times 10^9}
   {\sqrt{g_{\ast}} \: M_P \: J(x_{fr})} 
   \frac{1}{GeV}.  \label{omh2}
\end{eqnarray}
\noindent
The expansion of the annihilation cross section
$\sigma_{ann} v = a + b \: v^2$ is not a good approximation
in the vicinity of poles or thresholds \cite{griestv}.
In the present work the case of coannihilation is not important 
\cite{abeln, edsjoe}, since the LSP neutralino in the MSSM and NMSSM
is mostly bino-like or a mixed state with small mass 
(see Fig.\ref{bn}).
Furthermore the LSP singlinos \cite{ellws, ellwsi, king} of the NMSSM 
are now excluded by the UFB-3 constraint (Eq.(\ref{ufb3}))
as discussed in section 6.
\newline
\noindent
Compared to the MSSM the annihilation cross section of the neutralinos
changes in the NMSSM because of more particles in this model, 
which can mix, and modified vertices. 
In our numerical analysis we consider all possible decay channels.
The relevant formulas for the NMSSM are given in the Appendix A1.
They are obtained by substituting the MSSM couplings in \cite{drees} 
by the corresponding NMSSM couplings and by slightly modifying 
the partial wave amplitudes in \cite{drees}.

\section{Procedure of finding solutions}

\noindent
The SUSY soft breaking parameters $m_i$ and $A_i$ 
and the Yukawa couplings $\lambda$, $k$ and $h_t$    
are evolved with the SUSY RGEs (\cite{savoy}, Appendix A2) from the
GUT scale $M_X$ to the electroweak scale $M_{weak}$.
Since we assume universality at the GUT scale $M_X$ (Eq.(\ref{univ}))
we have only 6 GUT scale parameters $m_0$, $m_{1/2}$, 
$A_0$, $\lambda_0$, $k_0$ and $h_{t0}$.
Together with the low energy parameters $\tan \beta$ and $x$
there are 8 parameters in the NMSSM with only 5 of them
being independent.
In our procedure we take $\lambda_0$, $k_0$, $h_{t0}$, 
$\tan \beta$ and $x$ as independent input parameters.
The remaining 3 parameters ($m_0$, $m_{1/2}$, $A_0$)
are then determined by the three minimum conditions 
(\ref{nmin1}-\ref{nmin3}) and are numerically calculated 
by the method described in \cite{ellwtnb, ich}:
The three minimum conditions (\ref{nmin1}-\ref{nmin3}) depend 
only on the low energy parameters 
$m_i$, $A_i$, $\lambda$, $k$ and $h_t$    
generically denoted by $p_k(M_{weak})$: 
\begin{eqnarray}
   \frac{\partial \: V_{1 loop}}{\partial \: v_i} =
   F_i(p_k) = 0
   \: \: \: \: \: \: \: (i = 1-3).  \label{dvdv}
\end{eqnarray}
Since $\lambda_0$, $k_0$ and $h_{t0}$ are (randomly choosen)
fixed input parameters, all low energy parameters $p_k(M_{weak})$ 
are functions $f_k$ (which can be constant) of the GUT scale parameters
$m_0$, $m_{1/2}$ and $A_0$ only:
\begin{eqnarray}
   p_k(M_{weak}) = 
   f_k(m_0, m_{1/2}, A_0).
\end{eqnarray}
To obtain the functions $f_k$ we numerically solve the 
SUSY RGEs (\cite{savoy}, Appendix A2)
$dp_k \: / dt \: = \: h_k(t, p_1, ... , p_n) \: \: (k = 1 - n)$.
A solution to a system of first-order differential equations
can be found with the method of Runge, Kutta and Merson.
A subroutine for this numerical method is available in the 
CERN Program Library.
An additional subroutine has to be written which calculates 
the derivatives $dp_k \: / dt \: \: (k = 1 - n)$ depending on 
the current values of $t$ and $p_1$, ..., $p_n$.
In the three minimum conditions (\ref{dvdv}) we then substitute the 
low energy parameters $p_k(M_{weak})$ by the functions $f_k$
and obtain the generic equations
\begin{eqnarray}
   F_i(f_k(m_0, m_{1/2}, A_0)) = G_i(m_0, m_{1/2}, A_0) = 0
   \: \: \: \: \: \: \: (i = 1-3).  \label{fifk}
\end{eqnarray}
The SUSY breaking parameters $m_0$, $m_{1/2}$ and $A_0$
are then determined by the zero points of the generic equations
$G_i(m_0, m_{1/2}, A_0) = 0 \: \: (i = 1-3)$.
In order to find a solution of Eq.(\ref{fifk}) we first minimize
$\sum_{i=1}^{3} \: G_i(m_0, m_{1/2}, A_0)^2$ 
and then we check, whether the minimum is also a zero point.
The minimization is performed by using the simplex method 
of Nelder and Mead.
A corresponding minimization routine SIMPLEX is available
in the MINUIT package of the CERN Program Library.
The numerical procedure needs starting values for the parameters
$m_0$, $m_{1/2}$ and $A_0$, which are randomly generated
using a logarithmic measure within their ranges.
\newline
\noindent
To obtain a representative sample of solutions of the NMSSM,
we randomly generate $\sim 4.3 \times 10^8$ points in the
5 dimensional parameter space of the input parameters
$\lambda_0$, $k_0$, $h_{t0}$, $\tan \beta$ and $x$
using a logarithmic measure.
The SUSY breaking parameters $m_0$, $m_{1/2}$ and $A_0$ are then 
numerically calculated with the three minimum equations (\ref{fifk}).  
Having found a solution of the SUSY RGEs (\cite{savoy}, Appendix A2), 
whose effective 1 loop Higgs potential 
$V_{1 loop}^{Higgs}$ (Eq.(\ref{vh1l})) has a local minimum 
for the input parameters $\tan \beta$ and $x$, 
we impose the additional theoretical and experimental constraints 
described in section 3.
The very strong UFB-3 constraint (Eq.(\ref{ufb3})) excludes about 80\% 
of the solutions which fulfill all the other constraints of section 3.
For every solution we calculate the masses of all SUSY particles 
\cite{haber, bartl} and Higgses \cite{ellis, ellwp, elli2, king}.
Altogether we obtain about $6600$ solutions which fulfill 
all requirements.
The ranges of the parameters, for which we find solutions, 
are given by 
$m_0 \: \in \: [\: 60 \: GeV \: , \: 1500 \: GeV \: ]$,
$|m_{1/2}| \: \in \: [\: 120 \: GeV \: , \: 1500 \: GeV \: ]$,
$|A_0| \: \in \: [\: 180 \: GeV \: , \: 6000 \: GeV \: ]$,
$\lambda_0 \: \in \: [\: 2 \times 10^{-3} \: , \: 0.5 \: ]$,
$k_0 \: \in \: [\: 10^{-3} \: , \: 0.6 \: ]$,
$h_{t 0} \: \in \: [\: 0.4 \: , \: 0.9 \: ]$,
$|\tan \beta| \: \in \: [\: 2.4 \: , \: 20 \: ]$ and
$|x| \: \in \: [\: 800 \: GeV \: , \: 80000 \: GeV \: ]$.
Applying the dark matter constraint to the $6600$ solutions,
about $1500$ solutions remain which are cosmologically acceptable.

\section{Parameter space and particle spectra}

\begin{figure}
\vspace*{-3.cm}
\hspace*{-.5cm}
\epsfig{file=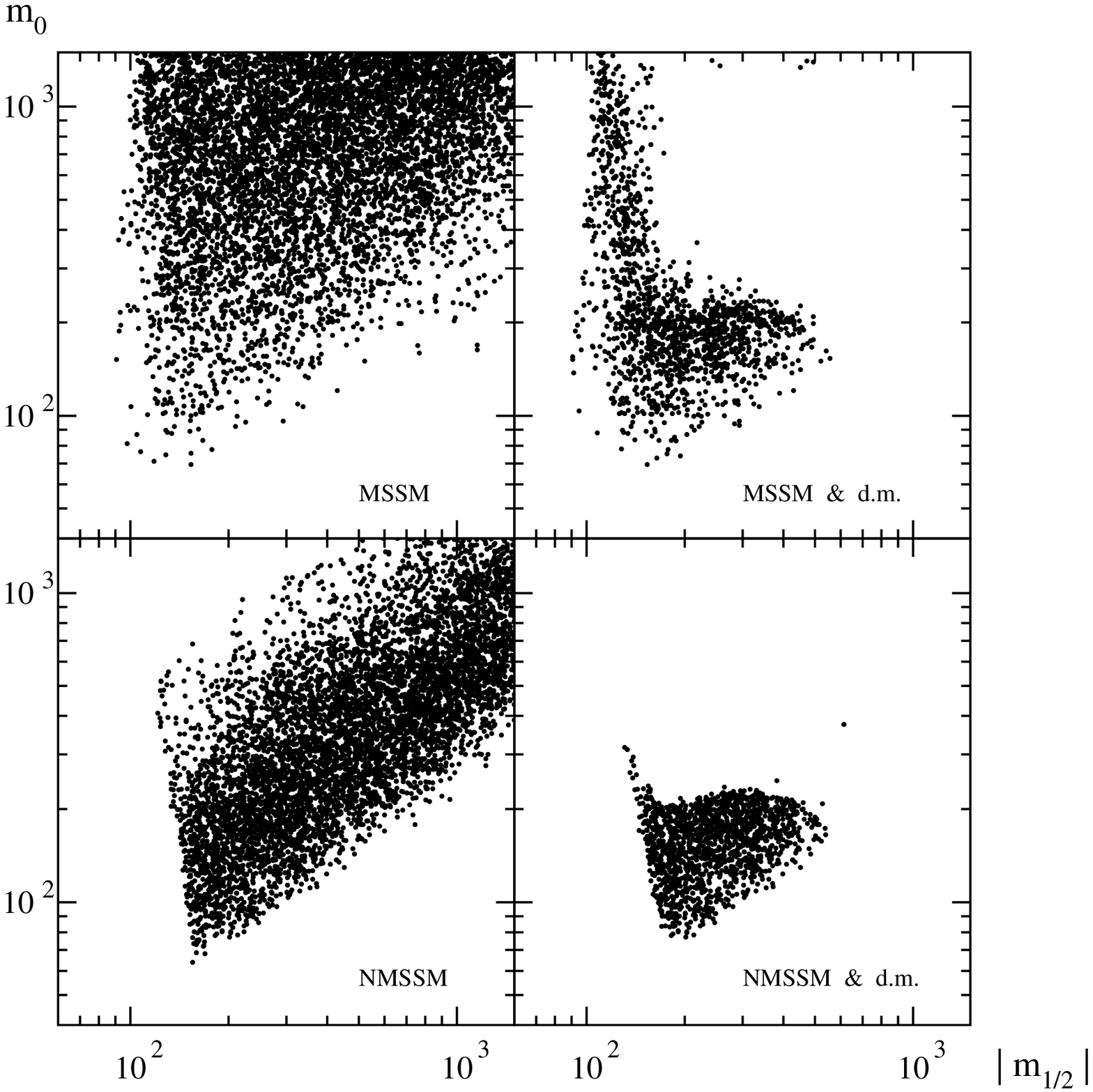,width=17cm}
\vspace*{.8cm}
\begin{center}
\refstepcounter{figure}
\label{mmg}
{\normalsize\bf Fig.\ \thefigure}
\end{center}
\end{figure}

\noindent
In the following we investigate the parameter space and the 
particle spectra of the MSSM and NMSSM.
The consequences of the dark matter condition for the two models 
are discussed.
We show our results in the form of scatter-plots.
Each point represents a solution of the SUSY RGEs 
(given in Appendix A2 for the NMSSM) with 
additional theoretical and experimental constraints imposed 
as described for the NMSSM in section 3.
In the upper left picture of Fig.\ref{mmg} the MSSM solutions 
are shown in the ($|m_{1/2}|$, $m_0$) plane.
The parameter region with small $m_0$ and large $|m_{1/2}|$
is excluded by the UFB-3 constraint \cite{casas}. 
Furthermore in the region $|m_{1/2}| \: \gg \: m_0$ 
the stau or stop would be the LSP.
But a charged or colored LSP like the stau or stop is 
cosmologically disfavored \cite{kane}.
In the upper right picture the cosmologically allowed 
solutions are shown.
The dark matter condition leads to an upper bound
of $600 \: GeV$ for $|m_{1/2}|$ and leaves $m_0$ unconstrained.
Values of $m_0$ larger than about $300 \: GeV$ are cosmologically allowed
due to Z-pole or Higgs-pole enhanced neutralino pair annihilation.
Comparable scatter-plots for the constrained MSSM 
are discussed in \cite{kolda}.
In that paper upper bounds of $940 \: GeV$ for $|m_{1/2}|$ 
and $500 \: GeV$ for $m_0$ are deduced for $m_t \: = \: 170 \: GeV$.
The lower left picture shows the NMSSM solutions.
The effect of the UFB-3 constraint (Eq.(\ref{ufb3})) in the NMSSM 
is comparable to the MSSM.
In the NMSSM this constraint also excludes the parameter region with 
small $m_0$ and large $|m_{1/2}|$.
The reason, why the case $m_0 \gg |m_{1/2}|$ is  
suppressed in the NMSSM, is discussed in \cite{ellws}.
It follows from the condition, that the minimum of the 
scalar potential does not break the conservation 
of charge and color,  
and from the condition that the physical minimum
is lower than the symmetric vacuum, $9 \: m_0^2 \: \lesssim \: A_0^2$.
In the lower right picture the dark matter condition 
is imposed leading to an upper bound of $650 \: GeV$ for $|m_{1/2}|$.
In contrast to the MSSM the upper bound for $m_0$ is about 
$300 \: GeV$ in the NMSSM.
These upper bounds follow from the cosmological upper bounds 
of $260 \: GeV$ for the LSP neutralino and $450 \: GeV$ for
the lighter selectron as discussed in \cite{ich} and below.
This can be understood from the fact \cite{drees}, that the 
LSP neutralino in the NMSSM is mostly bino-like (see Fig.\ref{bn}) 
and the annihilation cross section is dominated by sfermion exchange. 

\begin{figure}
\vspace*{-3.cm}
\hspace*{-.5cm}
\epsfig{file=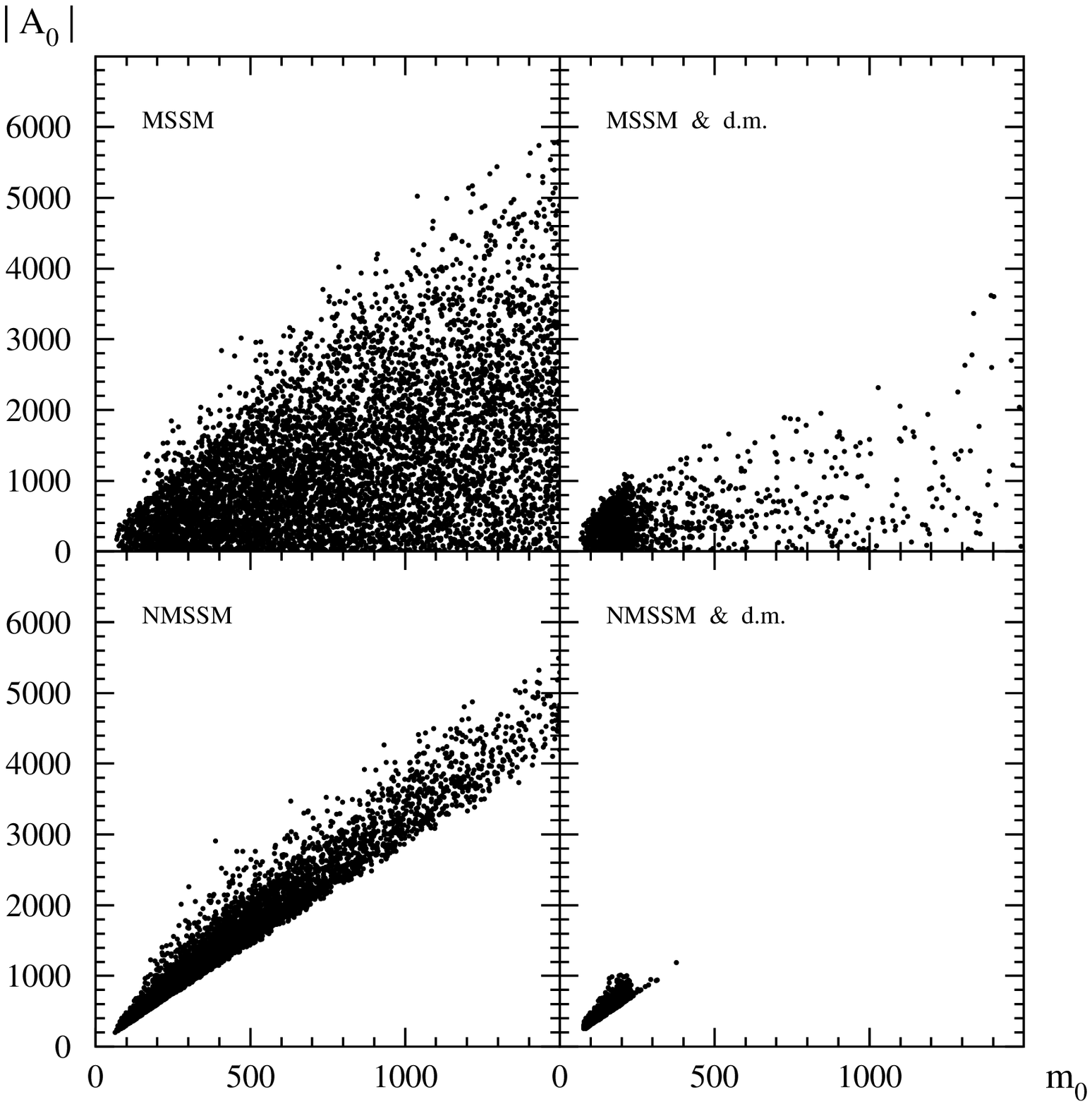,width=17cm}
\vspace*{.8cm}
\begin{center}
\refstepcounter{figure}
\label{ma}
{\normalsize\bf Fig.\ \thefigure}
\end{center}
\end{figure}

\noindent
For the MSSM we show in the upper left picture 
of Fig.\ref{ma} the solution points in the plane ($m_0$, $|A_0|$).
The upper bound for $|A_0|$ follows from the UFB-3 constraint \cite{casas}.
The upper right picture shows the cosmologically restricted 
solutions.
The concentration of points corresponds to $m_0$ values below
about $300 \: GeV$ in the upper right picture of Fig.\ref{mmg}.
Larger values of $m_0$ are cosmologically allowed because of 
Z-pole or Higgs-pole enhanced neutralino pair annihilation.
The lower left picture shows the allowed values of $|A_0|$ in the NMSSM.
Comparable with the MSSM the UFB-3 constraint (Eq.(\ref{ufb3}))
leads to an upper bound for $|A_0|$ in the NMSSM. 
In contrast to the MSSM there is also a lower bound for $|A_0|$
in the NMSSM.
The NMSSM solutions have to fulfill the relation
$9 \: m_0^2 \: \lesssim \: A_0^2$ \cite{ellws}, 
which follows from the condition that the physical minimum 
has to be lower than the symmetric vacuum. 
The dark matter condition strongly restricts the allowed range of $|A_0|$
as shown in the lower right picture.
Larger values of $m_0$ are cosmologically excluded,
since $m_0 \gg |m_{1/2}|$ is disfavored in the NMSSM 
as already discussed.

\begin{figure}
\vspace*{-3.cm}
\hspace*{-.5cm}
\epsfig{file=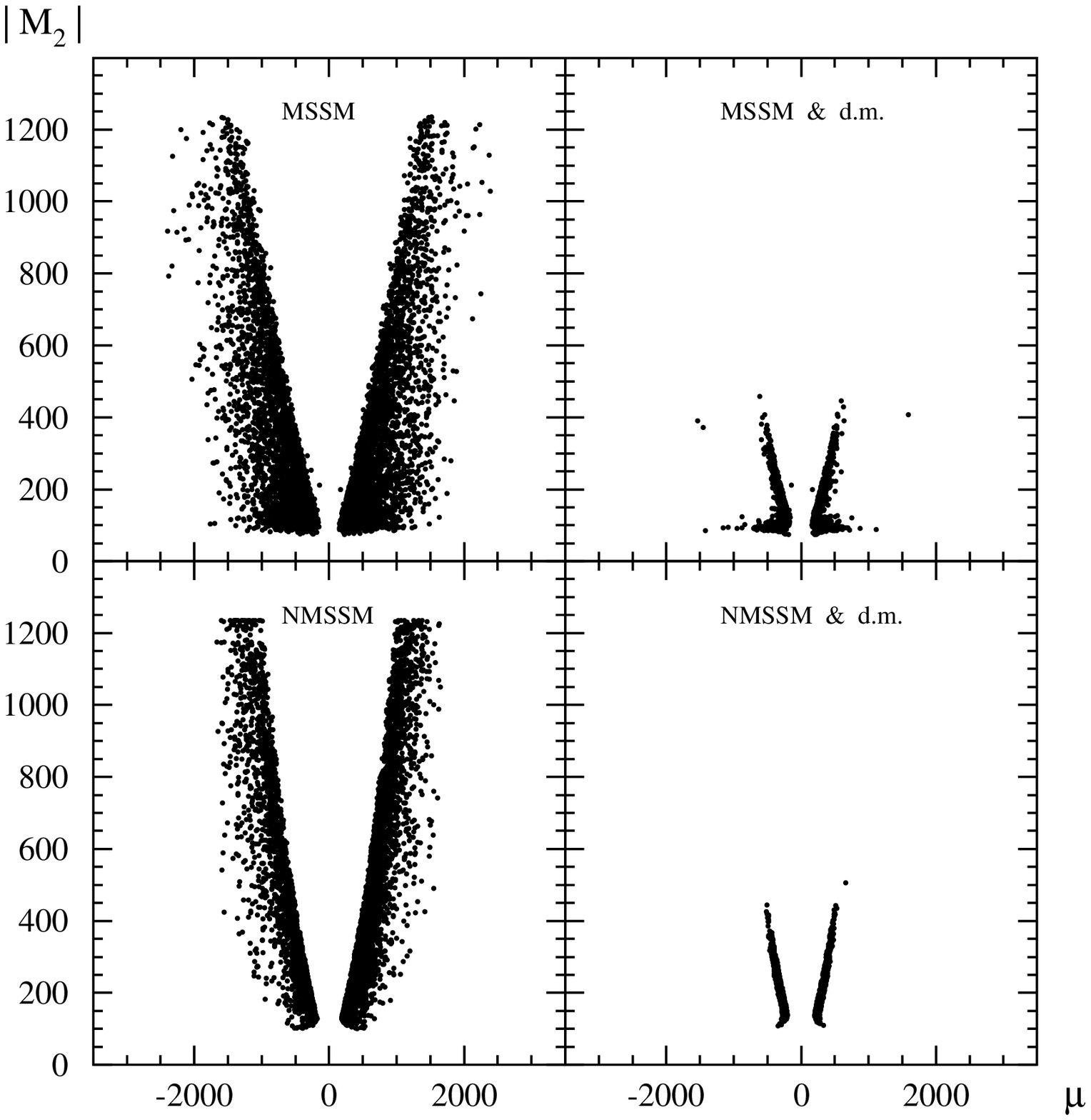,width=17cm}
\vspace*{.8cm}
\begin{center}
\refstepcounter{figure}
\label{mmu}
{\normalsize\bf Fig.\ \thefigure}
\end{center}
\end{figure}

\begin{figure}
\vspace*{-3.cm}
\hspace*{-.5cm}
\epsfig{file=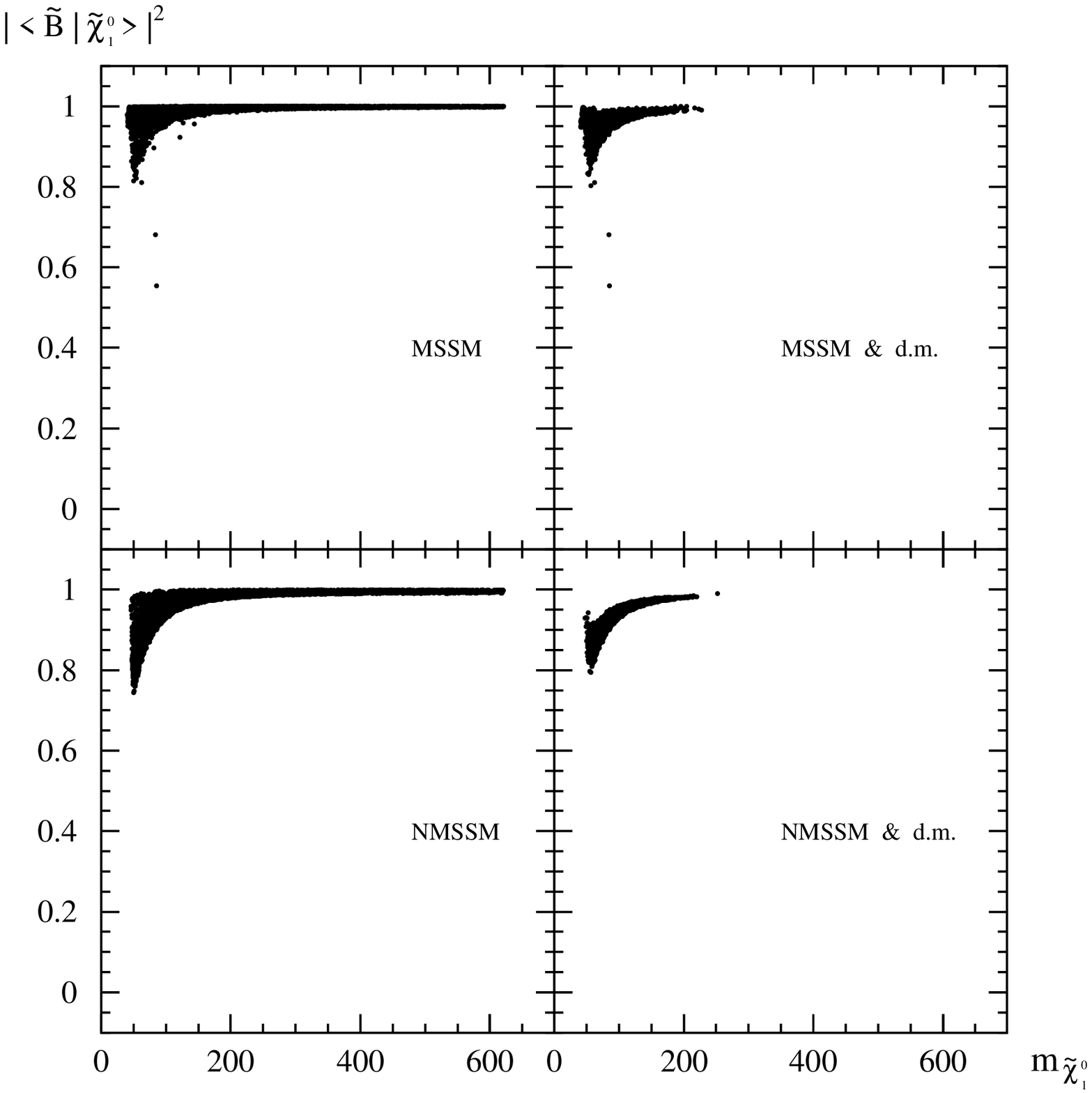,width=17cm}
\vspace*{.8cm}
\begin{center}
\refstepcounter{figure}
\label{bn}
{\normalsize\bf Fig.\ \thefigure}
\end{center}
\end{figure}

\noindent
In the upper left picture of Fig.\ref{mmu} we show the 
MSSM solution points in the plane ($\mu$, $|M_2|$).
The parameters $\mu$ and $M_2$ are taken at the 
electroweak scale $M_{weak}$.
The solutions are lying below the diagonals 
$|M_2| \: = \: |\mu |$.
The reason is the first minimum condition 
which favors larger values of $|\mu|$ for 
$|\tan \beta| \: \rightarrow \: 1$.
The first minimum condition of the MSSM has the same form as 
the first minimum condition in the NMSSM (Eq.(\ref{nmin1})) if $\mu$ is 
substituted for $\lambda \: x$.
This property of the solutions leads to bino-like LSP neutralinos 
in the MSSM (see Fig.\ref{bn}).
In the upper right picture the dark matter condition is imposed.
The upper bound of $500 \: GeV$ for $|M_2|$ corresponds to the 
upper bound of $600 \: GeV$ for $|m_{1/2}|$ in the
upper right picture of Fig.\ref{mmg}.
The NMSSM solutions in the lower left picture are similar to the 
MSSM solutions, but they are more restricted.
This restriction follows from the third minimum condition (Eq.(\ref{nmin3}))
in the NMSSM which would be an extra constraint for the MSSM.
The lower right picture shows the cosmologically allowed solutions.
The upper bound of $550 \: GeV$ for $|M_2|$ follows from the 
upper bound of $650 \: GeV$ for $|m_{1/2}|$ in the
lower right picture of Fig.\ref{mmg}.

\noindent
In the upper left picture of Fig.\ref{bn} the bino portion 
of the LSP neutralino $|<\tilde{B} \: | \: \tilde{\chi}^0_1 >|^2$
versus the LSP neutralino mass is shown for the MSSM.
The LSP neutralinos are mostly binos 
$|<\tilde{B} \: | \: \tilde{\chi}^0_1 >|^2 \, \approx \, 1$, 
which follows from the first minimum condition in the MSSM 
corresponding to Eq.(\ref{nmin1}) as already discussed.
The cosmologically allowed LSP neutralinos are shown in
the upper right picture. 
The dark matter condition sets an upper bound of $240 \: GeV$
for the LSP neutralino mass corresponding to the upper bound
of $600 \: GeV$ for $|m_{1/2}|$ in the upper right 
picture of Fig.\ref{mmg}.
The lower bound for the mass of the LSP neutralinos is about 
$40 \: GeV$ in the MSSM.
For the NMSSM the lower left picture shows the bino portion of 
the LSP neutralinos.
Similarly to the MSSM most of the LSP neutralinos are binos
in the NMSSM.
If the UFB-3 constraint (Eq.(\ref{ufb3})) is not applied, there exist
a large fraction of decoupled pure LSP singlinos with
$|<\tilde{B} \: | \: \tilde{\chi}^0_1 >|^2 \, = \, 0$
and a much smaller fraction of mixed LSP singlinos.
In \cite{ellwsi} the phenomenology of the LSP singlinos is investigated.
The LSP singlinos are now excluded 
by the UFB-3 constraint (Eq.(\ref{ufb3})), 
which excludes the parameter region with small $m_0$ 
and large $|m_{1/2}|$.
This is just the appropriate parameter region for LSP singlinos 
\cite{ellws, ellwsi, king}. 
The lower right picture shows the cosmologically allowed 
LSP neutralinos in the NMSSM.
The upper bound of $260 \: GeV$ for the LSP neutralino mass 
corresponds to the upper bound of $650 \: GeV$ for $|m_{1/2}|$ 
in the lower right picture of Fig.\ref{mmg}.
The dark matter condition improves the lower bound for  
the LSP neutralinos in the NMSSM to $50 \: GeV$ as 
discussed in \cite{ich}.
In the NMSSM all LSP neutralinos with mass 
$m_{\tilde{\chi}^0_1} \, \approx \, m_Z/2$ are 
cosmologically excluded, since their coupling to the Z boson 
is larger due to a larger Higgsino portion.

\begin{figure}
\vspace*{-3.cm}
\hspace*{-.5cm}
\epsfig{file=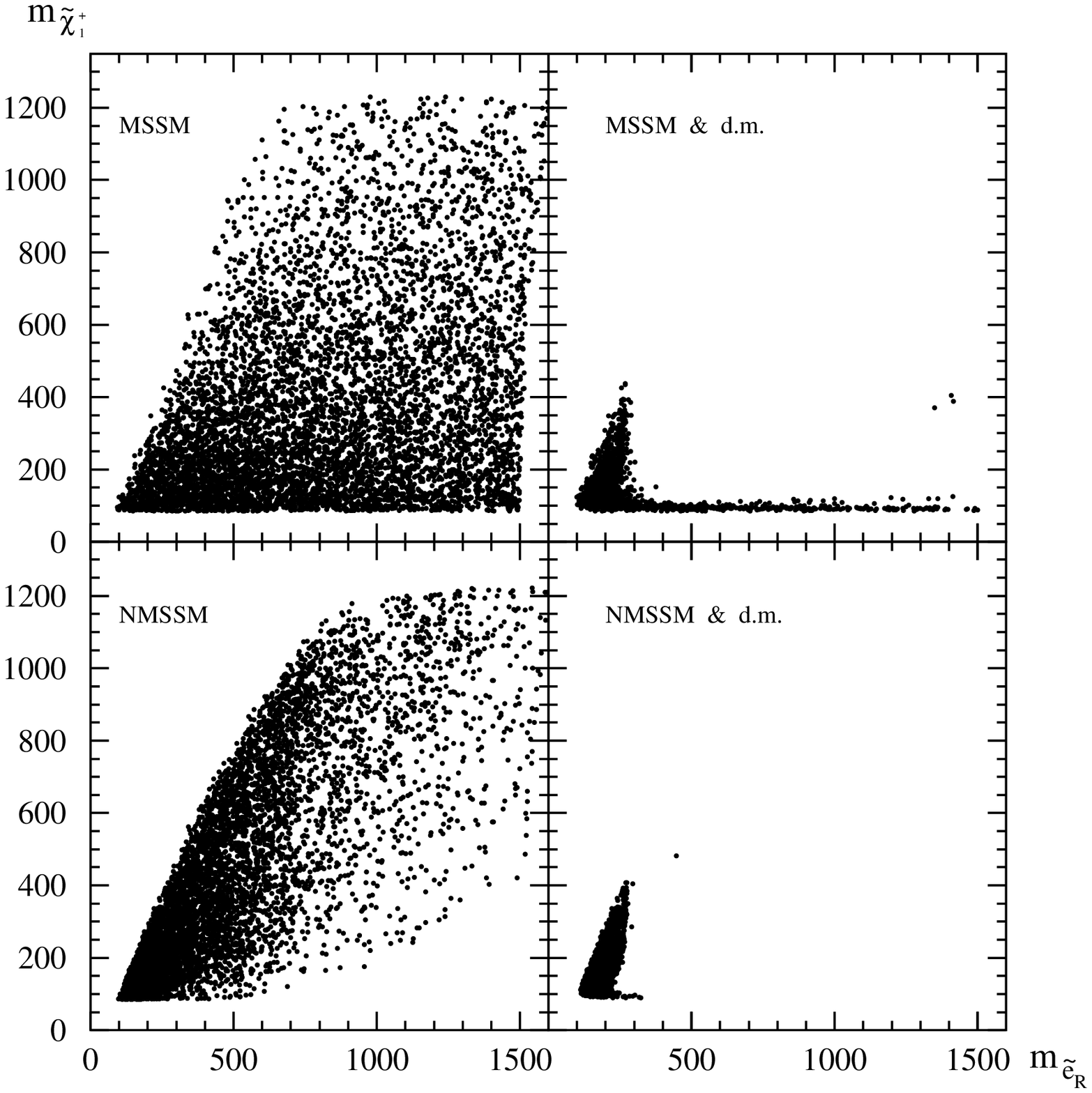,width=17cm}
\vspace*{.8cm}
\begin{center}
\refstepcounter{figure}
\label{ce}
{\normalsize\bf Fig.\ \thefigure}
\end{center}
\end{figure}

\noindent
In the upper left picture of Fig.\ref{ce} 
the mass of the lighter chargino versus the mass of
the lighter selectron is shown for the MSSM.
The upper right picture shows the cosmologically allowed
lighter chargino masses.
The dark matter condition sets un upper bound of $450 \: GeV$
for the lighter chargino and disfavors somewhat selectron masses 
larger than about $1 \: TeV$.
Selectrons with $300 \: GeV \: \lesssim \: m_{\tilde{e}_R}$
are cosmologically allowed due to Z-pole or Higgs-pole 
enhanced neutralino pair annihilation \cite{kolda}.
For the NMSSM the lower left picture shows the mass of the 
lighter chargino versus the mass of the lighter selectron.
Compared to the MSSM there are not so many solutions
in the NMSSM with small chargino mass and large selectron mass
following from the suppression of the domain
$m_0 \gg |m_{1/2}|$ in the NMSSM as discussed in \cite{ellws}. 
In the lower right picture the dark matter condition is imposed
leading to an upper bound of $500 \: GeV$ for the lighter 
chargino and a cosmologically allowed mass range of 
$100 \: GeV \: < \: m_{\tilde{e}_R} \: < \: 450 \: GeV$
for the lighter selectron in the NMSSM.
The cosmologically allowed solutions in the MSSM and NMSSM
show some differences.
Only in the MSSM there are cosmologically allowed heavier selectrons. 
In contrast to the MSSM, the dark matter condition leads to 
an slightly improved lower bound of $90 \: GeV$ for the 
lighter chargino in the NMSSM.
The reason is the enhanced lower bound of $50 \: GeV$ for 
the cosmologically allowed LSP neutralinos in the NMSSM 
as discussed and shown in Fig.\ref{bn} and 
the proportionality between the mass of the lighter chargino 
and the mass of the bino-like LSP neutralino.

\begin{figure}
\vspace*{-3.cm}
\hspace*{-.5cm}
\epsfig{file=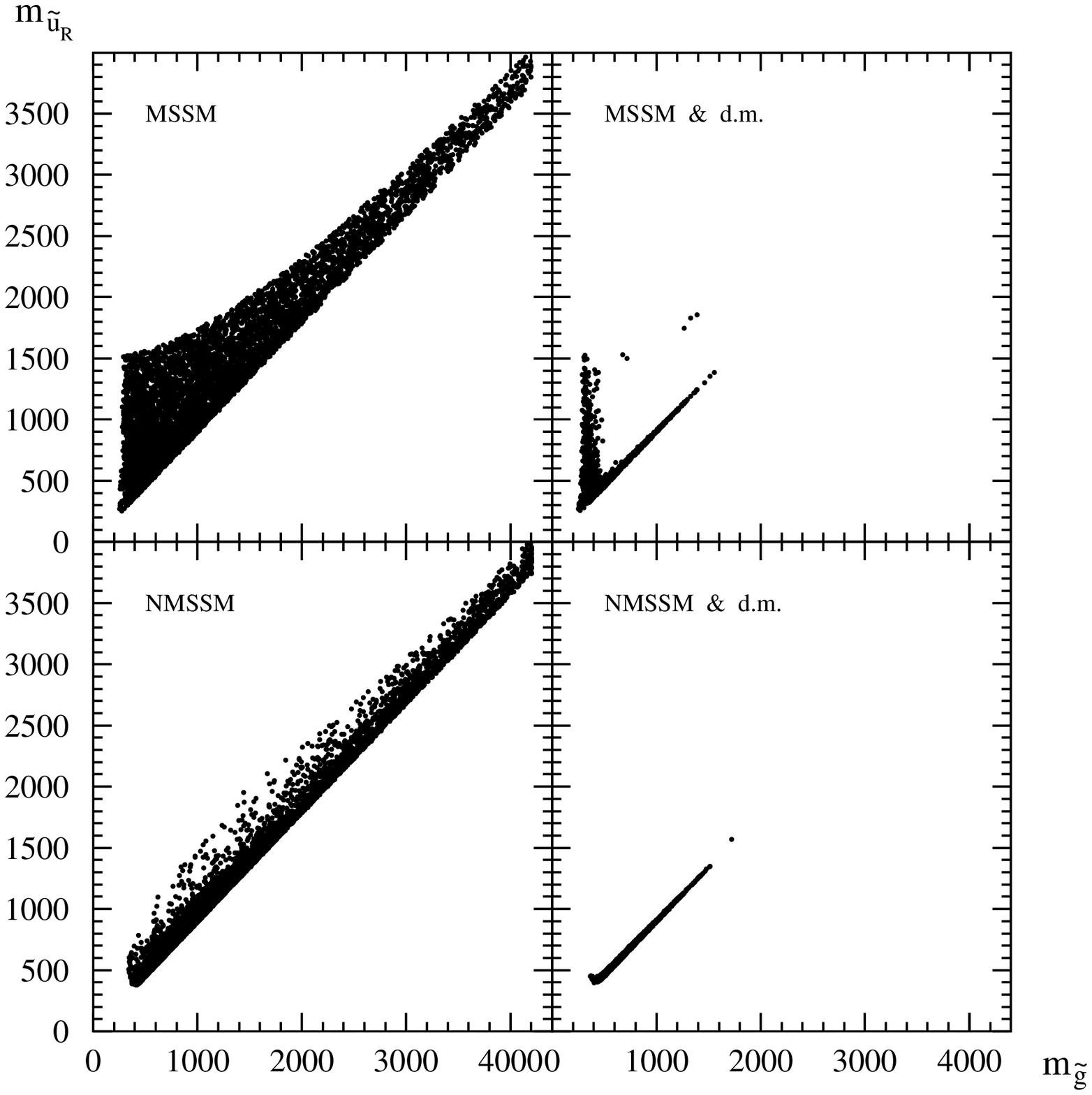,width=17cm}
\vspace*{.8cm}
\begin{center}
\refstepcounter{figure}
\label{ug}
{\normalsize\bf Fig.\ \thefigure}
\end{center}
\end{figure}

\noindent
In the upper left picture of Fig.\ref{ug} the mass of the lighter 
u-squark versus the gluino mass is shown for the MSSM.
The upper boundary line for the lighter u-squark mass results from the 
restriction of $m_0$. As mentioned in section 3 we have chosen $m_0$ 
to be smaller than $1500 \: GeV$ in order to avoid too much finetuning.
The lower bound for the gluino and the lighter u-squark is $250 \: GeV$.
In the upper right picture the cosmologically allowed 
lighter u-squark masses are shown.
In many cases the cosmologically allowed lighter u-squark mass
is proportional to the gluino mass.
The deviations from proportionality with large u-squark masses 
follow from the Z-poles or Higgs-poles shown in 
the upper right picture of Fig.1. 
The dark matter condition leads to an upper bound of $1600 \: GeV$ 
for the gluino following from the cosmological upper bound 
of $240 \: GeV$ for the bino-like LSP neutralino.
In the lower left picture the mass of the lighter u-squark 
is shown for the NMSSM.
In contrast to the MSSM \cite{barger}, 
the squark masses in the NMSSM
are nearly proportional to the gluino mass
as discussed in \cite{ich}.
The reason is, that in the NMSSM solutions with $m_0 \gg |m_{1/2}|$ 
are somewhat disfavored.
In contrast to the MSSM the lower bound for the gluino is $330 \: GeV$ 
in the NMSSM, which follows from the lower experimental bound of 
$85.4 \: GeV$ \cite{charg} for the chargino and from the lower bound of 
about $2.4$ for $|\tan \beta|$ in the NMSSM (see Fig.\ref{stb}) 
as discussed in \cite{ellws}.
Excluding small values of $m_0$ the UFB-3 constraint (Eq.(\ref{ufb3}))
further improves the lower bound for the lighter u-squark to $370 \: GeV$.
In the lower right picture the dark matter condition
is imposed.
In contrast to the MSSM, the dark matter condition improves
the lower bound for the gluino to $370 \: GeV$ in the NMSSM
as discussed in \cite{ich}.
The reason is the enhanced lower bound of $50 \: GeV$ and 
the larger Higgsino portion of the lightest cosmologically allowed 
LSP neutralinos in the NMSSM. 
With the dark matter condition the lower bound for the 
lighter u-squark slightly improves to $390 \: GeV$.
The cosmological upper bound of $1800 \: GeV$ for the gluino in the NMSSM
is similar to the corresponding upper bound in the MSSM. 
The upper bound of $1600 \: GeV$ for the lighter u-squark in the NMSSM
corresponds to the upper bound for the lighter u-squark in the MSSM
at large gluino masses. 

\begin{figure}
\vspace*{-3.cm}
\hspace*{-.5cm}
\epsfig{file=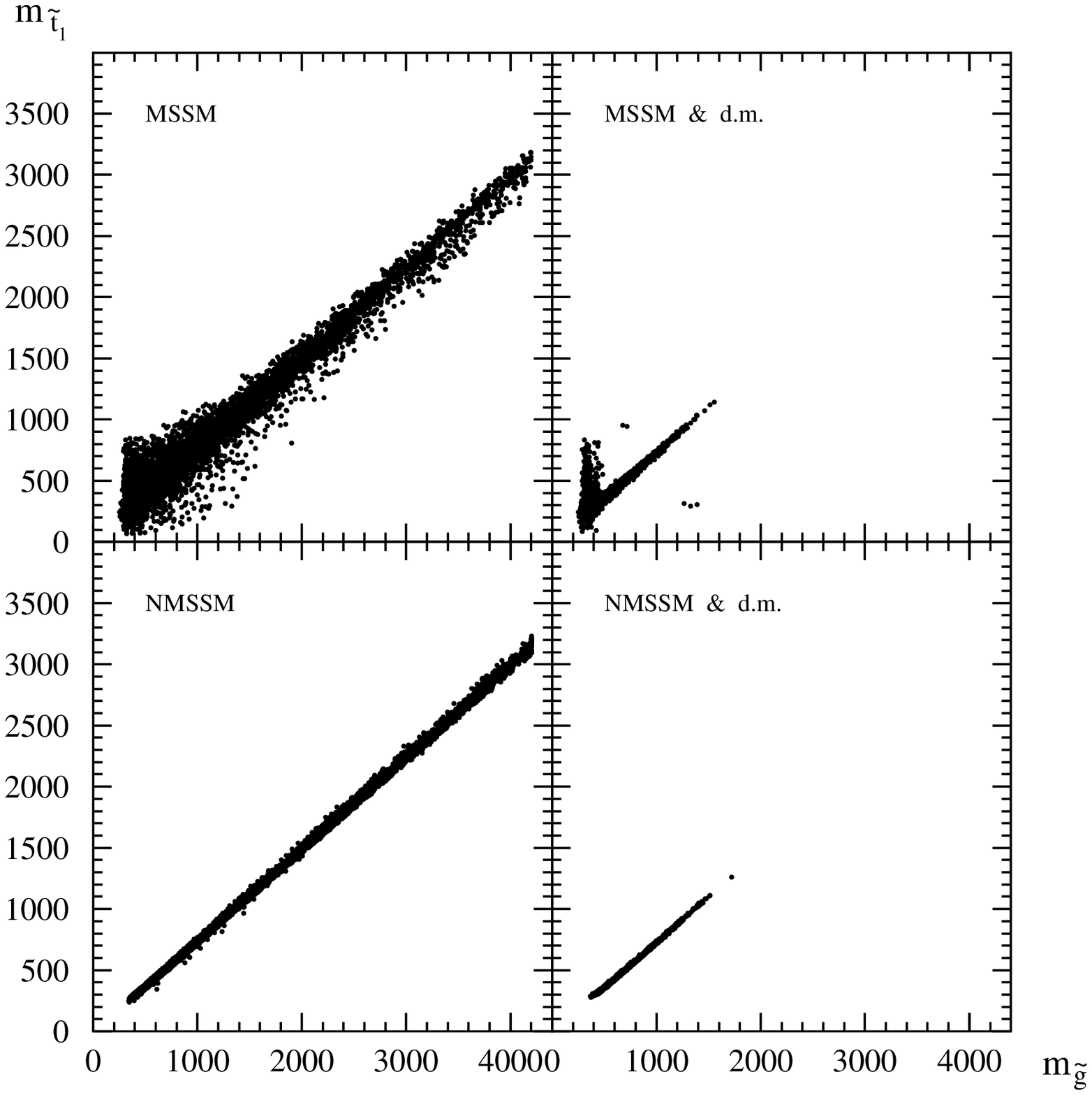,width=17cm}
\vspace*{.8cm}
\begin{center}
\refstepcounter{figure}
\label{tg}
{\normalsize\bf Fig.\ \thefigure}
\end{center}
\end{figure}

\begin{figure}
\vspace*{-3.cm}
\hspace*{-.5cm}
\epsfig{file=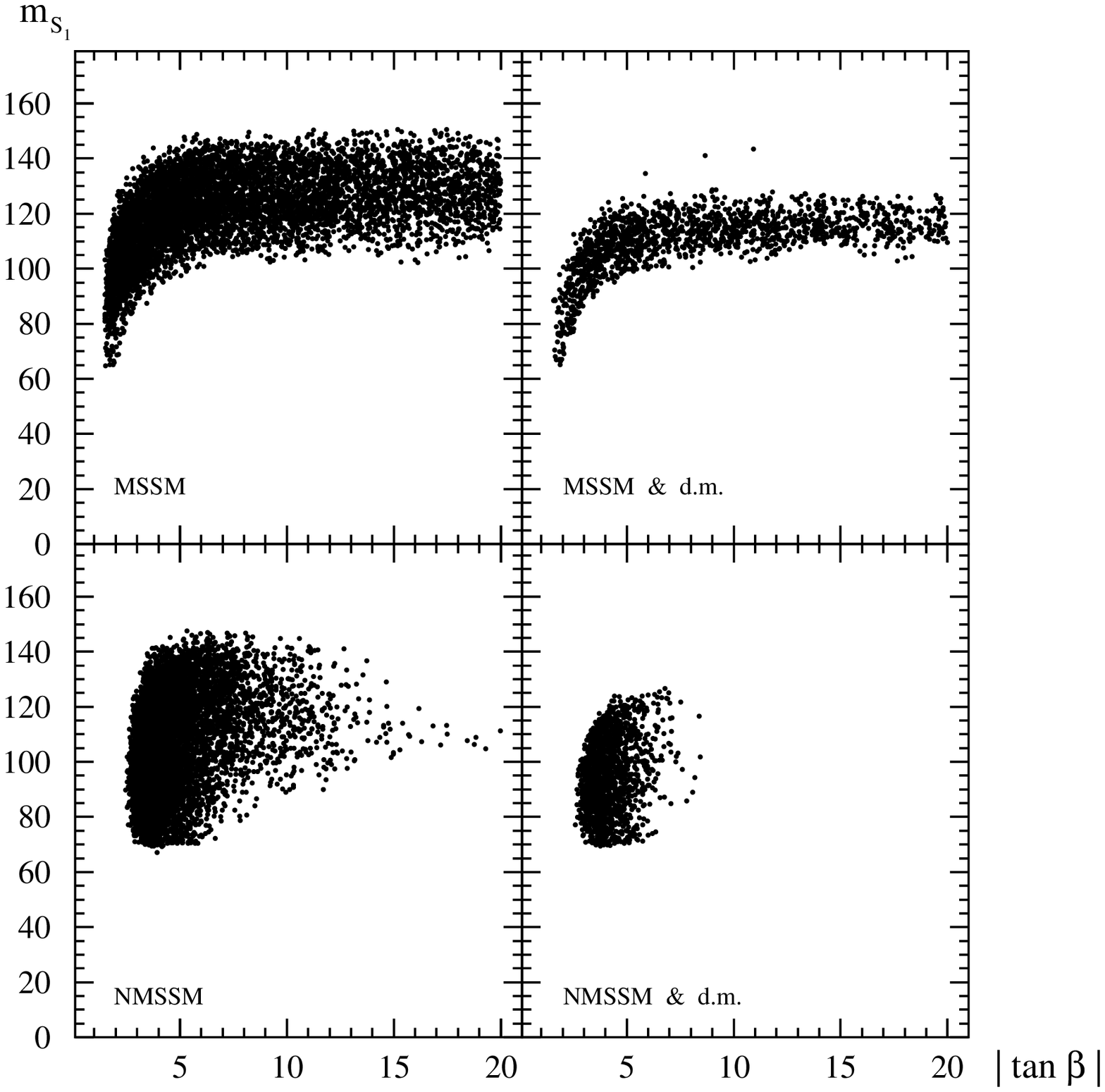,width=17cm}
\vspace*{.8cm}
\begin{center}
\refstepcounter{figure}
\label{stb}
{\normalsize\bf Fig.\ \thefigure}
\end{center}
\end{figure}

\noindent
In the upper left picture of Fig.\ref{tg} the mass of the lighter stop
versus the gluino mass is shown for the MSSM.
The stop masses are restricted from above, since the $m_0$ values
have been choosen to be smaller than $1500 \: GeV$
as discussed in section 3. 
In the upper right picture the cosmologically allowed 
lighter stop masses are shown.
Large stop masses at moderate gluino masses are allowed due to 
Z-poles and Higgs-poles.
The dark matter condition sets an upper bound of $1200 \: GeV$
for the lighter stop at large gluino masses.
In the lower left picture the lighter stop masses are shown
for the NMSSM.
In contrast to the MSSM \cite{barger}, 
the lighter stop mass is nearly proportional to the gluino mass 
as discussed in \cite{ich} following from the suppression of 
the domain $m_0 \gg |m_{1/2}|$ in the NMSSM.
In addition the constraints for the trilinear couplings
$A_{u_i}$, $A_{d_i}$ and $A_{e_i}$ 
(Eqs.(\ref{aui})-(\ref{aei})) are stronger in the NMSSM than in the MSSM.
The corresponding constraints in the MSSM are obtained 
by replacing $m^2_{H_{1,2}}$ in (Eqs.(\ref{aui})-(\ref{aei})) by 
$m^2_{H_{1,2}} + \mu^2$.
The stronger constraints for the trilinear couplings
lead to a larger mass of the stop in the NMSSM,
since the stop mixing depends on $A_t$.
Together with the UFB-3 constraint (Eq.(\ref{ufb3})), 
which excludes small values of $m_0$, the lower bound for the lighter stop 
increases to $240 \: GeV$ in the NMSSM.
The lower right picture shows the cosmologically restricted solutions.
The lower bound for the lighter stop slightly improves to $270 \: GeV$.
The cosmological upper bound of $1300 \: GeV$ for the lighter stop 
in the NMSSM is similar to the corresponding upper bound 
for the lighter stop in the MSSM at large gluino masses. 

\noindent
In the upper left picture of Fig.\ref{stb} we show the mass 
of the lighter scalar Higgs $S_1$ versus $|\tan \beta|$ for the MSSM.
In the MSSM the mass of the scalar Higgs $S_1$ has to be larger 
than the lower experimental bound of $64.5 \: GeV$ \cite{higgs}.
The upper bound for $m_{S_1}$ is $155 \: GeV$.
For $|\tan \beta|$ there is a lower bound of about $1.4$. 
In the upper right picture the dark matter condition is imposed
improving the upper bound for $m_{S_1}$ to about $145 \: GeV$.
The lower left picture shows the mass of the lightest 
scalar Higgs $S_1$ in the NMSSM, which is lower than $150 \: GeV$.
Before applying the UFB-3 constraint (Eq.(\ref{ufb3}))
there exist scalar Higgses $S_1$ in the NMSSM lighter than 
the MSSM Higgs bound of $64.5 \: GeV$ \cite{higgs}, 
which are predominantly a Higgs singlet $N$ \cite{ellwn, elli1, elli2}.
These very light scalar Higgses are now excluded by the 
UFB-3 constraint (Eq.(\ref{ufb3})), which excludes
the parameter region with small $m_0$ and large $|m_{1/2}|$.
But this parameter region is needed in order to have 
very light scalar Higgses with a large Higgs singlet portion
\cite{king}.
In contrast to the MSSM, the lower bound for $|\tan \beta|$
in the NMSSM is about $2.4$. 
The higher lower bound for $|\tan \beta|$ follows from
the extra third minimum condition (Eq.(\ref{nmin3})) in the NMSSM.
For larger values of $|\tan \beta|$ the density of NMSSM solutions
decreases, if the UFB-3 constraint (Eq.(\ref{ufb3})) is applied.
This follows from the fact that the NMSSM solutions with 
larger $|\tan \beta|$ are lying in the parameter region with 
small $m_0$ and large $|m_{1/2}|$ \cite{ellws, ellwsi}.
The lower right picture shows only solutions, 
which fulfill the dark matter condition. 
In the NMSSM the cosmologically allowed scalar Higgs $S_1$
is heavier than $69 \: GeV$ and lighter than $130 \: GeV$.
In contrast to the MSSM the dark matter condition 
together with the UFB-3 constraint (Eq.(\ref{ufb3})) leads to 
an upper bound of about 9 for $|\tan \beta|$ in the NMSSM.
However our investigations are restricted to 
$|\tan \beta|$ values below 20 as mentioned in section 3.

\begin{figure}
\vspace*{-3.cm}
\hspace*{-.5cm}
\epsfig{file=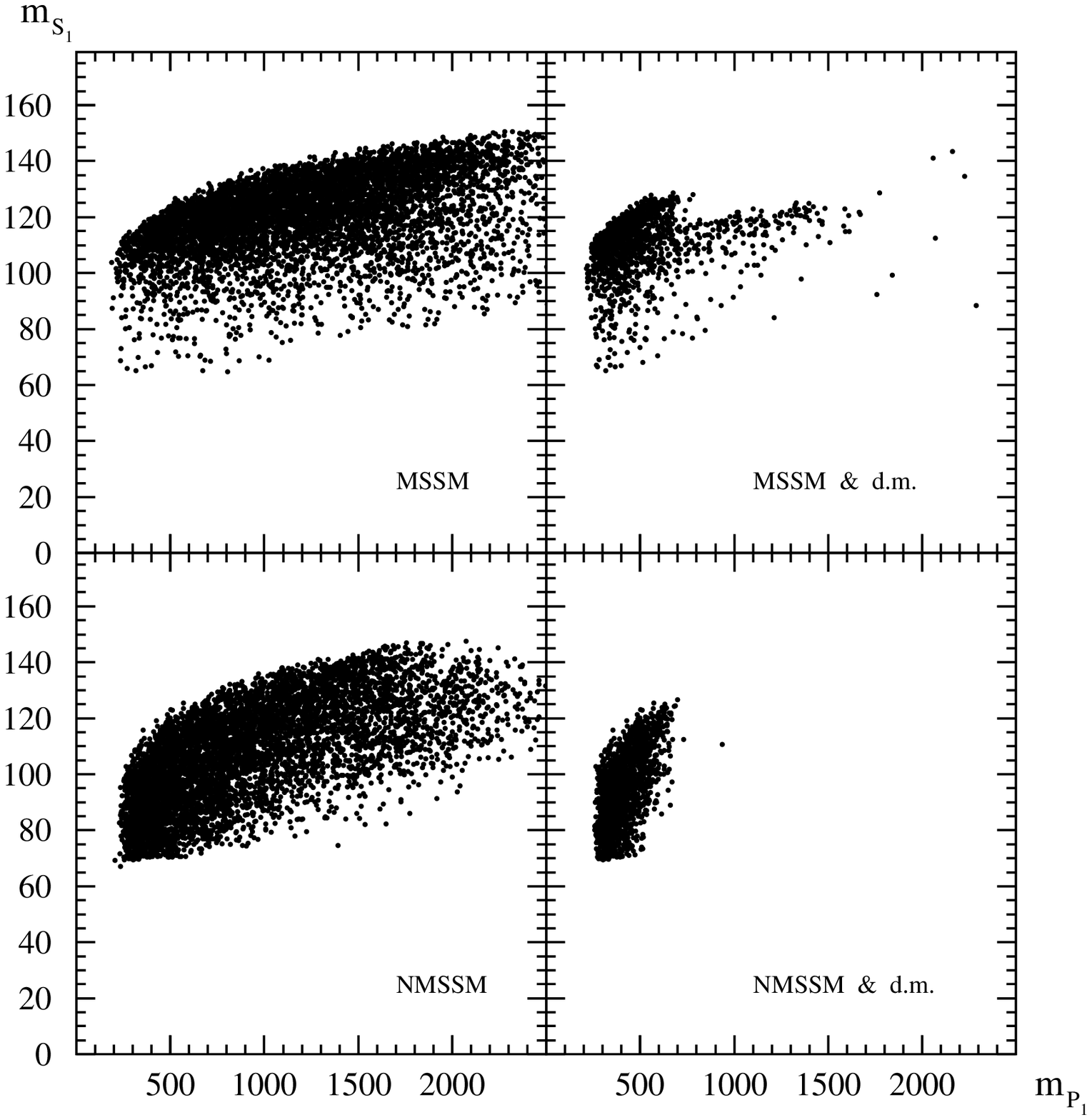,width=17cm}
\vspace*{.8cm}
\begin{center}
\refstepcounter{figure}
\label{ps}
{\normalsize\bf Fig.\ \thefigure}
\end{center}
\end{figure}

\noindent
In the upper left picture of Fig.\ref{ps} we show the mass 
of the lighter scalar Higgs $S_1$ versus the mass of the 
pseudoscalar Higgs $P_1$ for the MSSM.
In the MSSM $m_{P_1}$ is larger than $150 \: GeV$
within our restricted range of $|\tan \beta|$ values below 20
as mentioned in section 3. 
The upper right picture shows the cosmologically allowed Higgses.
The dark matter condition restricts the mass of the 
pseudoscalar Higgs $P_1$ to the range 
$200 \: GeV \: < \: m_{P_1} \: \lesssim \: 2300 \: GeV$.
The lower left picture shows the mass of the lightest scalar Higgs 
$S_1$ versus the mass of the lighter pseudoscalar Higgs 
$P_1$ for the NMSSM.
If the UFB-3 constraint (Eq.(\ref{ufb3})) is not applied
there exist very light scalar Higgses $S_1$ correlated with 
very light pseudoscalar Higgses $P_1$ which have both a large 
Higgs singlet portion \cite{ellwn}.
With the UFB-3 constraint (Eq.(\ref{ufb3})) these very light Higgses
are now excluded as already discussed 
for the very light scalar Higgs $S_1$ (see Fig.\ref{stb}).
For this reason we have now a lower bound of $200 \: GeV$
for the pseudoscalar Higgs $P_1$ in the NMSSM
for $|\tan \beta|$ values below 20.
In the lower right picture the dark matter condition is imposed
leading to a lower bound of $250 \: GeV$ and an upper bound of 
$900 \: GeV$ for the lighter pseudoscalar Higgs $P_1$.
Compared with the MSSM the upper bound for the lighter 
pseudoscalar Higgs $P_1$ is somewhat lower in the NMSSM.
This follows from the extra third minimum condition (Eq.(\ref{nmin3})) 
in the NMSSM and the suppression of the domain 
$m_0 \: \gg \: |m_{1/2}|$ in the NMSSM.

\section{Conclusions}

\noindent
In the present paper we have made a comparison between the MSSM 
and NMSSM concerning the parameter space and the particle spectra.
In contrast to the MSSM, the case $m_0 \: \gg \: |m_{1/2}|$
is disfavored in the NMSSM as discussed in \cite{ellws}
and the squark masses are nearly proportional to the gluino mass
(cf. Figs.(6) and (7)).
Furthermore the gluino and the lighter u-squark in the NMSSM 
are heavier than in the MSSM.
In the NMSSM the lower bound for the gluino is $330 \: GeV$ and 
$370 \: GeV$ for the lighter u-squark,
whereas the lower bound for the gluino and lighter u-squark
in the MSSM is $250 \: GeV$.
In the NMSSM a light stop is excluded by the stronger constraints 
for the trilinear couplings $A_{u_i}$, $A_{d_i}$ and $A_{e_i}$
(Eqs.(\ref{aui})-(\ref{aei})) and the UFB-3 constraint (Eq.(\ref{ufb3})). 
This leads to a larger stop mass 
$m_{\tilde{t}_1} \: > \: 240 \: GeV$ in the NMSSM
even before considering dark matter constraints.
This is in contrast to the MSSM where stops as light as
the experimental lower bound of $63 \: GeV$ \cite{sfer} are allowed.
One important difference between the NMSSM and the MSSM is the 
occurrence of LSP singlinos and light scalar and 
pseudoscalar Higgs singlets in the NMSSM.
Applying the UFB-3 constraint (Eq.(\ref{ufb3})) these singlet states 
are excluded as shown in Fig.4 and Fig.9 and discussed in section 6.
In this way the UFB-3 constraint strongly reduces the differences
between the NMSSM and the MSSM.
\newline
Furthermore we have investigated the consequences of the
dark matter condition for the MSSM and NMSSM.
Together with the UFB-3 constraint (Eq.(\ref{ufb3})) 
the dark matter condition gives an upper bound of about 9 
for $|\tan \beta|$ in the NMSSM (see Fig.8),
whereas in the MSSM larger values of $|\tan \beta|$ 
are furthermore allowed 
within our investigated range of $|\tan \beta|$ values below 20.
For the LSP neutralinos, lightest charginos, gluinos and 
scalar Higgses $S_1$ the cosmologically allowed mass ranges 
are comparable for the MSSM
($40 \: GeV  \: <  \: m_{\tilde{\chi}^0_1} \: < \: 240 \: GeV$,
$85.4 \: GeV  \: <  \: m_{\tilde{\chi}^+_1} \: < \: 450 \: GeV$,
$250 \: GeV \: < \: m_{\tilde{g}} \: < \: 1600 \: GeV$,
$64.5 \: GeV \: < \: m_{S_1} \: < \: 145 \: GeV$)
and for the NMSSM
($50 \: GeV  \: <  \: m_{\tilde{\chi}^0_1} \: < \: 260 \: GeV$,
$90 \: GeV  \: <  \: m_{\tilde{\chi}^+_1} \: < \: 500 \: GeV$,
$370 \: GeV \: < \: m_{\tilde{g}} \: < \: 1800 \: GeV$,
$69 \: GeV \: < \: m_{S_1} \: < \: 130 \: GeV$).
The enhanced lower bound of $50 \: GeV$ and the larger Higgsino portion 
of the lightest cosmologically allowed LSP neutralinos  
leads to improved lower bounds for the charginos and gluinos
in the NMSSM. 
Furthermore the dark matter condition slightly improves 
the lower bounds for the lighter u-squark and lighter stop
in the NMSSM to $390 \: GeV$ and $270 \: GeV$, respectively.
In the MSSM, due to Z-pole and Higgs-pole enhanced 
neutralino pair annihilation,
large masses of the selectrons, u-squarks and stops are
cosmologically allowed.
In contrast, the NMSSM does not allow for large slepton and
squark masses at moderate gluino masses (cf. Figs.(5-7))
and the dark matter condition leads to the upper bounds 
$m_{\tilde{e}_R} \: < \: 450 \: GeV$,
$m_{\tilde{u}_R} \: < \: 1600 \: GeV$ and   
$m_{\tilde{t}_1} \: < \: 1300 \: GeV$ at large gluino masses,
which are comparable with the MSSM.
In comparison with the MSSM the upper bound for the cosmologically 
allowed pseudoscalar Higgs $P_1$ is somewhat lower in the NMSSM 
$m_{P_1} \: < \: 900 \: GeV$ (Fig.9).
\newline
\noindent
Despite of the discussed differences, it may be difficult to 
distinguish experimentally between the MSSM and NMSSM.
One possibility is the measurement of smaller couplings 
of the neutral Higgses in the NMSSM,
which result from the mixing of the ordinary Higgs states 
with the Higgs singlets as discussed in \cite{ellwn, ellws}
and \cite{elli1} - \cite{king}. 
Another possibility suggested in \cite{ellws, ellwsi} is the search for 
an additional cascade in the NMSSM, which is produced in 
the decay of the bino-like second lightest neutralino
into the LSP singlino.
Since the LSP singlinos are excluded by the UFB-3 constraint
this additional cascade is not a way to discriminate
between the NMSSM and MSSM any more.

\section*{Acknowledgements}
\noindent
I am grateful to M. Gl\"uck and E. Reya 
for suggestions and many helpful discussions.
Also I would like to thank U. Ellwanger and C.A. Savoy
for useful conversations.
The work has been supported by the 
'Graduiertenkolleg am Institut f\"ur Physik
der Universit\"at Dortmund'.
\newpage
\section*{Appendix A1: Annihilation cross section}
\noindent
To calculate the relic neutralino density (Eq.(\ref{omh2})) in the NMSSM 
we need the annihilation cross section of the neutralinos 
into the different decay products $X$ and $Y$.
\[ \tilde{\chi}^0_1 (h) \: \: \tilde{\chi}^0_1 (\overline{h})
   \rightarrow X (\lambda_X) \: Y (\lambda_Y) \]
\noindent
$ \lambda_f = \lambda_X - \lambda_Y $, where $\lambda_X$ and $\lambda_Y$
are the helicities of particle $X$ and $Y$.
Within the partial wave formalism the annihilation cross section
is obtained \cite{drees} as:
\begin{eqnarray}
   \sigma_{ann}(\tilde{\chi}^0_1 \: \tilde{\chi}^0_1 
   \rightarrow X \: Y) v = \frac{1}{4} 
   \frac{\overline{\beta}_f}{8 \pi s S}
   \left[|A(^1S_0)|^2 + \frac{1}{3} 
   (|A(^3P_0)|^2 + |A(^3P_1)|^2 + |A(^3P_2)|^2 ) \right]  \label{siann}
\end{eqnarray}
\begin{eqnarray}
   \overline{\beta}_f = \sqrt{ 1 
   - \frac{2 (m_X^2 + m_Y^2)}{s} + \frac{(m_X^2 - m_Y^2)^2}{s^2} } 
\end{eqnarray}
\noindent
$s \approx 4 m_{\tilde{\chi}}^2$ is the center of mass energy squared.
S is a symmetry factor, which is 2 if $X = Y$. 
Summation over final helicities is implicit in Eq.(\ref{siann}).
The partial wave amplitude $A(^{2 S + 1}L_J)$ describes annihilation
from an initial state with spin S, orbital angular momentum L 
and total angular momentum J.
The partial wave amplitudes $A(^{2 S + 1}L_J)$ for the NMSSM are 
obtained by slightly modifying the partial wave amplitudes
of the MSSM in \cite{drees} and by using the corresponding
NMSSM couplings.
Thereby we have corrected some misprints \cite{dreesp} in \cite{drees}.
One modification of the neutralino annihilation is the exchange 
of 5 neutralinos in the NMSSM instead of 4 neutralinos in the MSSM.
Other modifications concern the exchange of the scalar Higgses 
$S_1$, $S_2$ and $S_3$ in the NMSSM 
instead of $H_1$ and $H_2$ in the MSSM and 
the exchange of the pseudoscalar Higgses 
$P_1$ and $P_2$ in the NMSSM 
instead of $P$ in the MSSM.
In addition the NMSSM coupling $g_{\alpha - +}$ (Eq.(\ref{gamp})) 
has to be inserted in Eq.(\ref{wha1s0}).
In the MSSM there is only one pseudoscalar Higgs $P$ and the 
coupling $g_{\alpha - +}$ turns out to be 1.  
In the following the partial wave amplitudes $A(^{2 S + 1}L_J)$ 
of the NMSSM for all decay channels are given in leading order
in the relative velocity $v$ of the non-relativistic neutralinos.
Previously we give some NMSSM couplings and definitions.
The couplings of the W boson to the chargino $\tilde{\chi}^+_j$
and the LSP neutralino, which is denoted with index 0 in \cite{drees}
instead of 1, have the same form in the MSSM and NMSSM,
but a different matrix $N$ (Eq.(\ref{xi})) enters:
\begin{eqnarray}
   O^L_{0 j} = - \frac{1}{\sqrt{2}} \, N_{0 4} \, V_{j 2}
   + N_{0 2} \, V_{j 1} 
\end{eqnarray}
\begin{eqnarray}
   O^R_{0 j} = \frac{1}{\sqrt{2}} \, N_{0 3} \, U_{j 2}
   + N_{0 2} \, U_{j 1}. 
\end{eqnarray}
\noindent
The mass matrix of the charginos is diagonalized by the orthogonal 
matrices $U$ and $V$, which are explicitly given in \cite{bartl}.
In the NMSSM the couplings of the charged Higgs to the 
chargino $\tilde{\chi}^+_j$ and the LSP neutralino 
has to be modified compared to the MSSM:
\begin{eqnarray}
   Q'^L_{0 j} = N_{0 4}\, V_{j 1} + \frac{1}{\sqrt{2}} \, 
   (N_{0 2} + N_{0 1} \, \tan \theta_W) \, V_{j 2}
   - \frac{\lambda \, \tan \beta}{g_2} \, N_{0 5} \, V_{j 2} 
\end{eqnarray}
\begin{eqnarray}
   Q'^R_{0 j} = N_{0 3}\, U_{j 1} - \frac{1}{\sqrt{2}} \, 
   (N_{0 2} + N_{0 1} \, \tan \theta_W) \, U_{j 2}
   - \frac{\lambda}{g_2 \, \tan \beta} \, N_{0 5} \, U_{j 2}. 
\end{eqnarray}
\noindent
In the MSSM and NMSSM the coupling of the Z boson to the neutralinos 
$\tilde{\chi}^0_i$ and $\tilde{\chi}^0_j$ is given by:
\begin{eqnarray}
   O''^L_{i j} = - \frac{1}{2} \, N_{i 3} \, N_{j 3}
   + \frac{1}{2} \, N_{i 4} \, N_{j 4}. 
\end{eqnarray}
\noindent
In the NMSSM the coupling of the scalar Higgs $S_a$ to the neutralinos 
$\tilde{\chi}^0_i$ and $\tilde{\chi}^0_j$ changes compared to the MSSM 
and can be written as:
\begin{eqnarray}
   T_{S_a i j} = - U_{a 1}^S \: Q''_{i j} + U_{a 2}^S \: S''_{i j}
   + U_{a 3}^S \: Z''_{i j} 
\end{eqnarray}
\begin{eqnarray}
   Q''_{i j} = \frac{1}{2 \: g_2} [ N_{i 3} \: 
   (g_2 \: N_{j 2} - g_y \: N_{j 1})
   + \sqrt{2} \: \lambda \: N_{i 4} \: N_{j 5} + (i \leftrightarrow j) ] 
\end{eqnarray}
\begin{eqnarray}
   S''_{i j} = \frac{1}{2 \: g_2} [ N_{i 4} \: 
   (g_2 \: N_{j 2} - g_y \: N_{j 1})
   - \sqrt{2} \: \lambda \: N_{i 3} \: N_{j 5} + (i \leftrightarrow j) ] 
\end{eqnarray}
\begin{eqnarray}
   Z''_{i j} = \frac{1}{2 \: g_2} 
   [ - \sqrt{2} \: \lambda \: N_{i 3} \: N_{j 4} 
   + \sqrt{2} \: k \: N_{i 5} \: N_{j 5} + (i \leftrightarrow j) ]. 
\end{eqnarray}
\noindent
The $3 \times 3$ mass matrix of the scalar Higgses $S_i$ \cite{ellis} 
in the basis ($H^0_{1 R}$, $H^0_{2 R}$, $N_R$) with 
$H^0_{i R} \: = \: \sqrt{2} \: Re\{H^0_i\} \:\: (i = 1 - 3)$
is diagonalised by the matrix $U^S$.
The scalar Higgses $S_i \:\: (i = 1 - 3)$ are ordered
with increasing mass. 
The coupling of the pseudoscalar Higgs $P_{\alpha}$ 
to the neutralinos is also different in the MSSM and NMSSM.
In the NMSSM this coupling has the following form:
\begin{eqnarray}
   T_{P_{\alpha} i j} = - U_{\alpha 1}^P \: Q'''_{i j} 
   + U_{\alpha 2}^P \: S'''_{i j} - U_{\alpha 3}^P \: Z'''_{i j} 
\end{eqnarray}
\begin{eqnarray}
   Q'''_{i j} = \frac{1}{2 \: g_2}  
   [ N_{i 3} \: (g_2 \: N_{j 2} - g_y \: N_{j 1})
   - \sqrt{2} \: \lambda \: N_{i 4} \: N_{j 5} + (i \leftrightarrow j) ] 
\end{eqnarray}
\begin{eqnarray}
   S'''_{i j} = \frac{1}{2 \: g_2} 
   [ N_{i 4} \: (g_2 \: N_{j 2} - g_y \: N_{j 1})
   + \sqrt{2} \: \lambda \: N_{i 3} \: N_{j 5} + (i \leftrightarrow j) ] 
\end{eqnarray}
\begin{eqnarray}
   Z'''_{i j} = Z''_{i j}.  
\end{eqnarray}
\noindent
The $3 \times 3$ mass matrix of the pseudoscalar Higgses $P_{\alpha}$ 
and the neutral Goldstone boson \cite{ellis} in the basis 
($H^0_{1 I}$, $H^0_{2 I}$, $N_I$) with 
$H^0_{i I} \: = \: \sqrt{2} \: Im\{H^0_i\} \:\: (i = 1 - 3)$
is here diagonalised by the $3 \times 3$ matrix $U^P$.
The pseudoscalar Higgses $P_{\alpha} \:\: (i = 1, 2)$ are ordered
with increasing mass. 
In \cite{drees} the following definitions are used:
\begin{eqnarray}
  \Delta^2 = \frac{m^2_X + m^2_Y}{2 \, m^2_{\tilde{\chi}}} 
\end{eqnarray}
\begin{eqnarray}
  R_X = \frac{m_X}{m_{\tilde{\chi}}}, \:\: etc.  
\end{eqnarray}
\begin{eqnarray}
  P_I = 1 + R_I^2 - (R_X^2 + R^2_Y)/2.
\end{eqnarray}

\[ 1. \:\:\: \tilde{\chi}^0_1 \: \tilde{\chi}^0_1 \rightarrow 
   W^-(\lambda) \: W^+(\overline{\lambda}). \]
\noindent
The W bosons are produced by Z boson and scalar Higgs 
$S_i \: (i = 1, 2, 3)$ exchange in the s-channel and chargino 
$\tilde{\chi}^+_j \: (j = 1,2)$ exchange in the t- and u-channel. 

\vspace*{.2cm}
\noindent
$ A(^1S_0): \lambda_f = 0, \lambda = \pm 1 $
\[ 2 \, \sqrt{2} \, \overline{\beta}_f \, g^2_2 \,
   \frac{O^{L 2}_{0 j} + O^{R 2}_{0 j}}{P_j} \]
\begin{eqnarray}
   + \sqrt{2} \, v^2 \, \overline{\beta}_f \, g^2_2 \,
   \left[ \frac{2}{3} \, \frac{R^+_j}{P^2_j} \, 
   O^L_{0 j} \, O^R_{0 j}
   + \frac{O^{L 2}_{0 j} + O^{R 2}_{0 j}}{P_j} \, 
   \left[ \frac{1}{4} - \frac{4}{3 \, P_j}  
   + \frac{2 \, \overline{\beta}_f^2}{3 \, P^2_j} \right] \right] 
\end{eqnarray}

\noindent
$ A(^3P_0): \lambda_f = 0, \lambda = 0 $
\[ \frac{\sqrt{6} \, v \, g^2_2}{R^2_W} \,
   \left[ - \frac{4}{3} \, \frac{O^{L 2}_{0 j} + O^{R 2}_{0 j}}{P_j}
   + \frac{4 \, O^L_{0 j} \, O^R_{0 j} \, R^+_j}{P_j} \,
   \left[ 1 - \frac{2}{3 \, P_j} \right] \right] \]
\[ + \, \sqrt{6} \, v \, g^2_2 \,
   \left[ \frac{O^{L 2}_{0 j} + O^{R 2}_{0 j}}{P_j} \,
   \left[ 1 - \frac{2 \, \overline{\beta}_f^2}{3 \, P_j} \right] \,
   - \frac{2 \, O^L_{0 j} \, O^R_{0 j} \, R^+_j}{P_j} \,
   \left[ 1 - \frac{4}{3 \, P_j} \right] \right] \]
\begin{eqnarray}
   - \, \frac{\sqrt{6} \, v \, (1 + \overline{\beta}_f^2) \, g^2_2 \, F_i}
   {(4 - R^2_{S_i} + i G_{S_i}) \, R_W} 
\end{eqnarray}

\noindent
$ \lambda_f = 0, \lambda = \pm 1 $
\begin{eqnarray}
   \sqrt{6} \, v \, g^2_2 \,
   \left[ \frac{O^{L 2}_{0 j} + O^{R 2}_{0 j}}{P_j} \,
   \left[ \frac{1}{3} - \frac{2 \, \overline{\beta}_f^2}{3 \, P_j} 
   \right] \,
   - \frac{2 \, O^L_{0 j} \, O^R_{0 j} \, R^+_j}{P_j} \, \right] 
   + \, \frac{\sqrt{6} \, v \, g^2_2 \, R_W \, F_i}
   {4 - R^2_{S_i} + i G_{S_i}} 
\end{eqnarray}

\noindent
$ A(^3P_1): | \lambda_f | = 1 $
\[ \frac{2 \, v \, \overline{\beta}_f^2 \, \lambda_f \, g^2_2}{R_W}
   \left[ \frac{O^{L 2}_{0 j} + O^{R 2}_{0 j}}{P_j} \, 
   \left[ 1 - \frac{1}{P_j} \right] 
   - \frac{2 \, O^L_{0 j} \, O^R_{0 j} \, R^+_j}{P^2_j} \, \right] \]
\begin{eqnarray}
   + 2 \, v \, \overline{\beta}_f \, g^2_2 \,
   \frac{O^{L 2}_{0 j} - O^{R 2}_{0 j}}{R_W \, P_j} \,
   \left[ 2 - \frac{\overline{\beta}_f^2}{P_j} \right]  
   - \, \frac{8 \, v \, \overline{\beta}_f \, g^2_2 \, O''^L_{00}}
   {R_W \, (4 - R^2_Z)}  
\end{eqnarray}

\noindent
$ \lambda_f = 0, \lambda = 0 $
\begin{eqnarray}
   \frac{2 \, v \, \overline{\beta}_f}{R^2_W \, P_j} \,
   (3 - \overline{\beta}_f^2) \, g^2_2 \, 
   (O^{L 2}_{0 j} - O^{R 2}_{0 j}) 
   - \frac{4 \, v \, \overline{\beta}_f \, g^2_2 \, O''^L_{00}}
   {(4 - R^2_Z) \, R^2_W} \, (3 - \overline{\beta}_f^2)  
\end{eqnarray}

\noindent
$ \lambda_f = 0, \lambda = \pm 1 $
\begin{eqnarray}
   \frac{2 \, v \, \overline{\beta}_f}{P_j} \, g^2_2 \, 
   (O^{L 2}_{0 j} - O^{R 2}_{0 j}) 
   - \frac{4 \, v \, \overline{\beta}_f \, g^2_2 \, O''^L_{00}}
   {4 - R^2_Z}  
\end{eqnarray}

\noindent
$ A(^3P_2): | \lambda_f | = 2 $
\begin{eqnarray}
   - \, \frac{2 \, \sqrt{2} \, v}{P_j} \, g^2_2 \,
   (O^{L 2}_{0 j} + O^{R 2}_{0 j}) 
\end{eqnarray}

\noindent
$ | \lambda_f | = 1 $
\begin{eqnarray}
   \frac{2 \, v \, g^2_2}{R_W} \,
   \left[ \frac{- R^{+ 2}_j}{P^2_j} \,  
   (O^{L 2}_{0 j} + O^{R 2}_{0 j}) \,
   + \frac{2 \, O^L_{0 j} \, O^R_{0 j} \, R^+_j}{P^2_j} \, 
   \overline{\beta}_f^2 + \lambda_f \, \overline{\beta}_f^{\, 3} \, 
   \frac{O^{L 2}_{0 j} - O^{R 2}_{0 j}}{P^2_j} \right]     
\end{eqnarray}

\noindent
$ \lambda_f = 0, \lambda = \pm 1 $
\begin{eqnarray}
   \frac{2 \, v \, g^2_2}{\sqrt{3}} \, (O^{L 2}_{0 j} + O^{R 2}_{0 j}) \,
   \frac{1 - R^{+ 2}_j - R^2_W}{P^2_j} 
\end{eqnarray}

\noindent
$ \lambda_f = 0, \lambda = 0 $
\begin{eqnarray}
   \frac{4 \, v \, g^2_2}{\sqrt{3} \, R^2_W} \, 
   \left[ - \frac{O^{L 2}_{0 j} + O^{R 2}_{0 j}}{P_j} \,
   \left[ 1 - \overline{\beta}_f^2 \, \frac{R^2_W}{P_j} \right] 
   + 4 \, \overline{\beta}_f^2 \,
   \frac{O^L_{0 j} \, O^R_{0 j} \, R^+_j}{P^2_j} \right]  
\end{eqnarray}

\noindent
$G_{S_i} \: = \: \Gamma_{S_i} \: m_{S_i} \: / \: m_{\tilde{\chi}}^2$
are the rescaled widths of the scalar Higgses $S_i$.
The coupling of the scalar Higgs $S_i$ to the LSP neutralinos
and to the W bosons is combined to $F_i$.
In the NMSSM $F_i$ is given by: 
\begin{eqnarray}
   F_i = (U^S_{i 1} \, \cos \beta + U^S_{i 2} \, \sin \beta) \, 
   T_{S_i 0 0}. 
\end{eqnarray}

\[ 2. \:\:\: \tilde{\chi}^0_1 \: \tilde{\chi}^0_1 \rightarrow 
   Z(\lambda) \: Z(\overline{\lambda}). \]
\noindent
The Z bosons are produced by scalar Higgs $S_i \: (i = 1, 2, 3)$
exchange in the s-channel and neutralino $\tilde{\chi}^0_j \:  (j = 1-5)$
exchange in the t- and u-channel. 

\vspace*{.2cm}
\noindent
$ A(^1S_0): \lambda_f = 0, \lambda = \pm 1 $
\begin{eqnarray}
   \frac{4 \, \sqrt{2} \, \overline{\beta}_f \, g^2_2 \, 
   O''^{L 2}_{0 j}}{\cos^2 \theta_W \, P_j}
   + \frac{2 \, \sqrt{2} \, v^2 \, \overline{\beta}_f \, g^2_2 \, 
   O''^{L 2}_{0 j}}{\cos^2 \theta_W} \,
   \left[ - \frac{R_j}{3 \, P^2_j} + \frac{1}{P_j} \, 
   \left[ \frac{1}{4} - \frac{4}{3 \, P_j}  
   + \frac{2 \, \overline{\beta}_f^2}{3 \, P^2_j} \right] \right] 
\end{eqnarray}

\noindent
$ A(^3P_0): \lambda_f = 0, \lambda = 0 $
\[ \frac{4 \, \sqrt{6} \, v \, g^2_2}{\cos^2 \theta_W \, R^2_Z} \,
   O''^{L 2}_{0 j} \, \left[ - \frac{2}{3 \, P_j} - \frac{R_j}{P_j} \,
   \left[ 1 - \frac{2}{3 \, P_j} \right] \right] \]
\begin{eqnarray}
   + \, \frac{2 \, \sqrt{6} \, v \, g^2_2 \, O''^{L 2}_{0 j}}
   {\cos^2 \theta_W} \, \left[ \frac{1}{P_j} \,
   \left[ 1 - \frac{2 \, \overline{\beta}_f^2}{3 \, P_j} \right] \,
   + \frac{R_j}{P_j} \, \left[ 1 - \frac{4}{3 \, P_j} \right] \right] 
   - \frac{\sqrt{6} \, v \, g^2_2 \, (1 + \overline{\beta}_f^2) \, F_i}
   {(4 - R^2_{S_i} + i G_{S_i}) \, R_W} 
\end{eqnarray}

\noindent
$ \lambda_f = 0, \lambda = \pm 1 $
\begin{eqnarray}
   \frac{2 \, \sqrt{6} \, v \, g^2_2 \, O''^{L 2}_{0 j}}{\cos^2 \theta_W} \,
   \left[ \frac{1}{P_j} \, \left[ \frac{1}{3} 
   - \frac{2 \, \overline{\beta}_f^2}{3 \, P_j} \right] \,
   + \frac{R_j}{P_j} \, \right] 
   + \, \frac{\sqrt{6} \, v \, g^2_2 \, R_W \, F_i}
   {\cos^2 \theta_W \, (4 - R^2_{S_i} + i G_{S_i})} 
\end{eqnarray}

\noindent
$ A(^3P_1): | \lambda_f | = 1 $
\begin{eqnarray}
   \frac{4 \, v \, \overline{\beta}_f^2 \, \lambda_f \, g^2_2 \,
   O''^{L 2}_{0 j}}{\cos^2 \theta_W \, R_Z}
   \left[ \frac{1}{P_j} \, \left[ 1 - \frac{1}{P_j} \right] 
   + \frac{R_j}{P^2_j} \, \right] 
\end{eqnarray}
 
\noindent
$ A(^3P_2): | \lambda_f | = 2 $
\begin{eqnarray}
   - \, \frac{4 \, \sqrt{2} \, v \, g^2_2}{\cos^2 \theta_W \, P_j} \,
   O''^{L 2}_{0 j} 
\end{eqnarray}

\noindent
$ | \lambda_f | = 1 $
\begin{eqnarray}
   - \, \frac{4 \, v \, g^2_2}{\cos^2 \theta_W \, R_Z} \,
   O''^{L 2}_{0 j} \, \left[ \frac{R^2_j}{P^2_j} \,  
   + \frac{R_j}{P^2_j} \, \overline{\beta}_f^2 \right]     
\end{eqnarray}

\noindent
$ \lambda_f = 0, \lambda = \pm 1 $
\begin{eqnarray}
   \frac{4 \, v \, g^2_2}{\sqrt{3} \, \cos^2 \theta_W} \, 
   O''^{L 2}_{0 j} \, \frac{1 - R^2_j - R^2_Z}{P^2_j} 
\end{eqnarray}

\noindent
$ \lambda_f = 0, \lambda = 0 $
\begin{eqnarray}
   - \, \frac{8 \, v \, g^2_2}{\sqrt{3} \, \cos^2 \theta_W \, R^2_Z} \, 
   O''^{L 2}_{0 j} \, \left[ \frac{1}{P_j} \,
   \left[ 1 - \frac{R^2_Z \, \overline{\beta}_f^2}{P_j} \right] 
   + \frac{2 \, R_j \, \overline{\beta}_f^2}{P^2_j} \right]  
\end{eqnarray}

\[ 3. \:\:\: \tilde{\chi}^0_1 \: \tilde{\chi}^0_1 \rightarrow 
   Z(\lambda) \: S_a. \]
\noindent
The Z boson and the scalar Higgs $S_a$ are produced by 
Z boson and pseudoscalar Higgs $P_{\alpha} \: (\alpha = 1, 2)$
exchange in the s-channel and 
neutralino $\tilde{\chi}^0_j \:  (j = 1-5)$
exchange in the t- and u-channel. 

\vspace*{.2cm}
\noindent
$ A(^1S_0): \lambda = 0 $
\[ - \, \frac{2 \, \sqrt{2} \: \overline{\beta}_f}{R_Z} \,
   \frac{g^2_2}{\cos \theta_W} \, \left[ 
   \frac{2 \, J_j \, (R_j - 1)}{P_j} 
   + \frac{O}{R_Z \, \cos \theta_W} 
   - \frac{4 \, L_{\alpha}}{4 - R^2_{P_{\alpha}} + i G_{P_{\alpha}}} 
   \right] \]
\[ - \, v^2 \, \frac{\sqrt{2} \: \overline{\beta}_f}{R_Z} \,
   \frac{g^2_2}{\cos \theta_W} \, \frac{J_j}{P_j} \, \left[ 
   \frac{1}{2} \, (R_j - 5) - \frac{2 \, (R_j - 1)}{P_j} 
   + \frac{4 \, (R_j - 1)}{3 \, P^2_j} \, \overline{\beta}_f^2 
   + (2 - \Delta^2) \, \frac{2}{3 \, P_j} \right] \]
\begin{eqnarray}
   + \, v^2 \, \frac{\sqrt{2} \: \overline{\beta}_f}{R_Z} \,
   \frac{g^2_2}{\cos \theta_W} \, \left[
   \frac{3 \, L_{\alpha}}{4 - R^2_{P_{\alpha}} + i G_{P_{\alpha}}}
   - \frac{O}{4 \, R_Z \, \cos \theta_W} \right] 
\end{eqnarray}

\noindent
$ A(^3P_1): \lambda = \pm 1 $
\begin{eqnarray}
   - 4 \, v \, \frac{g^2_2}{\cos \theta_W} \, \frac{J_j}{P^2_j} \, 
   \left[ R^2_j - \frac{\delta^4}{4} \right]
   - 4 \, v \, \frac{g^2_2}{\cos \theta_W} \, \frac{J_j \, R_j}{P_j}
   + 2 \, v \, \frac{g^2_2}{\cos^2 \theta_W} \, O \,
   \frac{R_Z}{4 - R^2_Z} 
\end{eqnarray}

\noindent
$ \lambda = 0 $
\begin{eqnarray}
   - 2 \, \frac{v}{R_Z} \, \left[ 1 + \frac{\delta^2}{2} \right] \, 
   \frac{g^2_2}{\cos \theta_W} \, \left[
   2 \, (1 + R_j) \, \frac{J_j}{P_j}
   - \frac{R_Z}{\cos \theta_W \, (4 - R^2_Z)} \, O \right] 
\end{eqnarray}

\noindent
$ A(^3P_2): \lambda = \pm 1 $
\begin{eqnarray}
   - 4 \, \lambda \, v \, \overline{\beta}_f^2 \, 
   \frac{g^2_2}{\cos \theta_W} \, 
   \frac{J_j}{P_j^2}  
\end{eqnarray}
\noindent
In \cite{drees} the following definition is used:
\begin{eqnarray}
   \delta^2 = (R^2_Z - R^2_{S_a})/2. 
\end{eqnarray}
\noindent
$G_{P_{\alpha}} \: = \: \Gamma_{P_{\alpha}} \: m_{P_{\alpha}} \: / \: 
m_{\tilde{\chi}}^2$ are the rescaled widths of the pseudoscalar Higgses 
$P_{\alpha}$.
The couplings appearing in the neutralino $\tilde{\chi}^0_j$ 
exchange are combined to $J_j$: 
\begin{eqnarray}
   J_j = - O''^L_{0 j} \, T_{S_a 0 j}. 
\end{eqnarray}
\noindent
$L_{\alpha}$ is the product of the coupling of the pseudoscalar Higgs 
$P_{\alpha}$ to the LSP neutralinos and to the $Z$ boson and the
scalar Higgs $S_a$.
In the NMSSM $L_{\alpha}$ reads:  
\begin{eqnarray}
   L_{\alpha} = \frac{1}{2} \, 
   (U^S_{a 2} \, U^P_{\alpha 2} - U^S_{a 1} \, U^P_{\alpha 1}) \, 
   T_{P_{\alpha} 0 0}.  
\end{eqnarray}
\noindent
$O$ consists of the couplings appearing in the Z boson exchange.
In the NMSSM $O$ is given by:  
\begin{eqnarray}
   O = O''^L_{0 0} \, 
   (U^S_{a 1} \, \cos \beta + U^S_{a 2} \, \sin \beta). 
\end{eqnarray}
\newpage

\[ 4. \:\:\: \tilde{\chi}^0_1 \: \tilde{\chi}^0_1 \rightarrow 
   Z(\lambda) \: P_{\alpha}. \]
\noindent
The Z boson and the pseudoscalar Higgs $P_{\alpha}$
are produced by scalar Higgs $S_i \: (i = 1, 2, 3)$
exchange in the s-channel and neutralino 
$\tilde{\chi}^0_j \:  (j = 1-5)$ exchange in the t- and u-channel. 

\vspace*{.2cm}
\noindent
$ A(^3P_0): \lambda = 0 $
\[ 4 \, \sqrt{6} \, \frac{v \, \overline{\beta}_f}{R_Z} \,
   \frac{g^2_2}{\cos \theta_W} \, \left[ 
   \left[ 1 + \frac{\delta^2}{2} \right]
   \frac{K_j \, R_j}{3 \, P^2_j} \right. \]
\begin{eqnarray}
   \left. - \left[ \frac{2}{3} + R^2_j - \frac{2 \, \Delta^2}{3} 
   + \frac{\delta^2}{6} \right] \, \frac{K_j}{P^2_j}
   + \frac{L_i}{4 - R^2_{S_i}} \right] 
\end{eqnarray}

\noindent
$ A(^3P_1): \lambda = \pm 1 $
\begin{eqnarray}
   - 4 \, v \, \overline{\beta}_f \, \lambda \, 
   \frac{g^2_2}{\cos \theta_W} \, 
   \left[ - R_j + \frac{\delta^2}{2} \right]
   \frac{K_j}{P^2_j} 
\end{eqnarray}

\noindent
$ A(^3P_2): \lambda = \pm 1 $
\begin{eqnarray}
   4 \, v \, \overline{\beta}_f \, \frac{g^2_2}{\cos \theta_W} \, 
   \left[ - R_j + \frac{\delta^2}{2} \right] \frac{K_j}{P^2_j} 
\end{eqnarray}

\noindent
$ \lambda = 0 $
\begin{eqnarray}
   - \frac{8}{\sqrt{3}} \, v \, \overline{\beta}_f \,  
   \frac{g^2_2}{\cos \theta_W} \, \frac{K_j}{R_Z \, P^2_j} 
   \left[ 1 + R_j - \Delta^2 + \frac{\delta^2}{2} \, 
   (R_j -1) \right] 
\end{eqnarray}

\noindent
In \cite{drees} the following definition is used:
\begin{eqnarray}
   \delta^2 = (R^2_Z - R^2_{P_{\alpha}})/2. 
\end{eqnarray}
\noindent
The couplings appearing in the neutralino $\tilde{\chi}^0_j$ exchange 
are combined to $K_j$: 
\begin{eqnarray}
   K_j = O''^L_{0 j} \, T_{P_{\alpha} 0 j}. 
\end{eqnarray}
\noindent
$L_i$ is the product of the coupling of the scalar Higgs $S_i$
to the LSP neutralinos and to the $Z$ boson and the 
pseudoscalar Higgs $P_{\alpha}$.
In the NMSSM $L_i$ has the following form:
\begin{eqnarray}
   L_i = \frac{1}{2} \, 
   (U^S_{i 2} \, U^P_{\alpha 2} - U^S_{i 1} \, U^P_{\alpha 1}) \, 
   T_{S_i 0 0}.  
\end{eqnarray}

\[ 5. \:\:\: \tilde{\chi}^0_1 \: \tilde{\chi}^0_1 \rightarrow 
   W^-(\lambda) \: H^+. \]
\noindent
The W boson and charged Higgs are produced by scalar Higgs 
$S_i \: (i = 1, 2, 3)$ and pseudoscalar Higgs 
$P_{\alpha} \: (\alpha = 1, 2)$ exchange in the s-channel 
and chargino $\tilde{\chi}^+_j \: (j = 1,2)$
exchange in the t- and u-channel. 

\vspace*{.2cm}
\noindent
$ A(^1S_0): \lambda = 0 $
\[ 4 \, \sqrt{2} \, \overline{\beta}_f \, g^2_2 \, 
   \frac{J'_j + J''_j}{R_W \, P_j} + \sqrt{2} \, v^2 \, 
   \frac{\overline{\beta}_f \, g^2_2 \, (J'_j + J''_j)}{R_W \, P_j}
   \left[ \frac{5}{2} - \frac{2}{P_j} 
   + \frac{4 \, \overline{\beta}_f^2}{3 \, P^2_j}
   - \frac{2}{3 \, P_j} \, (2 - \Delta^2) \right] \]
\[ - \, 4 \, \sqrt{2} \, \overline{\beta}_f \, g^2_2 \, 
   (J'_j - J''_j) \, \frac{R^+_j}{R_W \, P_j} 
   - \, \sqrt{2} \, \overline{\beta}_f \, v^2 \, g^2_2 \, 
   (J'_j - J''_j) \, \frac{R^+_j}{R_W \, P_j} 
   \left[ \frac{4 \, \overline{\beta}_f^2}{3 \, P^2_j}
   - \frac{2}{P_j} + \frac{1}{2} \right] \]
\begin{eqnarray}
   - \, 4 \, \sqrt{2} \, \overline{\beta}_f \, g^2_2 \, 
   g_{\alpha - +} \, T_{P_{\alpha} 0 0} \, \frac{1}{R_W} \, 
   \frac{1}{4 - R^2_{P_{\alpha}}} \, 
   \left[ 1 + \frac{3 \, v^2}{8} \right]  \label{wha1s0}
\end{eqnarray}

\noindent
$ A(^3P_0): \lambda = 0 $
\[ 4 \, \sqrt{6} \, \frac{v \, \overline{\beta}_f}{R_W} \, g^2_2
   \left[ R^+_j \left[1 + \frac{\delta^2}{2} \right] \,
   \frac{K'_j - K''_j}{3 \, P^2_j}
   - \left[ \frac{2}{3} + R^{+ 2}_j - \frac{2}{3} \, \Delta^2 
   + \frac{\delta^2}{6} \right] \, \frac{K'_j + K''_j}{P^2_j} \right] \]
\begin{eqnarray}
   - \, \frac{4 \, \sqrt{6}}{R_W} \, v \, \overline{\beta}_f \,
   \frac{g^2_2 \, L_i}{4 - R^2_{S_i}} 
\end{eqnarray}

\noindent
$ A(^3P_1): \lambda = \pm 1 $
\[ - \, 4 \, v \, \left[ R^{+ 2}_j - \frac{\delta^4}{4} \right] \,
   g^2_2 \, \frac{J'_j + J''_j}{P^2_j}
   - \, 4 \, v \, g^2_2 \, (J'_j - J''_j) \, \frac{R^+_j}{P_j} \]
\begin{eqnarray}
   + \, \frac{4 \, v \, \overline{\beta}_f \, \lambda}{P^2_j} \,
   g^2_2 \, \left[ R^+_j \, (K'_j - K''_j)
   - \frac{\delta^2}{2} \, (K'_j + K''_j) \right] 
\end{eqnarray}

\noindent
$ \lambda = 0 $
\begin{eqnarray}
   - \, \frac{4 \, v \, g^2_2}{R_W \, P_j} \, \left[ 
   (J'_j + J''_j) + (J'_j - J''_j) \, R^+_j \right]
   \left[1 + \frac{\delta^2}{2} \right] 
\end{eqnarray}

\noindent
$ A(^3P_2): \lambda = \pm 1 $
\begin{eqnarray}
   - \, 4 \, v \, \lambda \, g^2_2 \,
   \frac{J'_j + J''_j}{P^2_j} \, \overline{\beta}_f^2
   + \frac{4 \, v \, \overline{\beta}_f}{P^2_j} \,
   g^2_2 \, \left[ (K'_j + K''_j) \, \frac{\delta^2}{2} 
   - R^+_j \, (K'_j - K''_j) \right] 
\end{eqnarray}

\noindent
$ \lambda = 0 $
\begin{eqnarray}
   \frac{8 \, v \, \overline{\beta}_f \, g^2_2}{\sqrt{3} \, R_W \, P^2_j} 
   \, \left[ \left[ -1 + \Delta^2 + \frac{\delta^2}{2} \right]
   \, (K'_j + K''_j) - R^+_j \, \left[ 1 + \frac{\delta^2}{2} \right] \,
   (K'_j - K''_j) \right] 
\end{eqnarray}

\noindent
In \cite{drees} the following definition is used:
\begin{eqnarray}
   \delta^2 = (R^2_W - R^2_{H^+})/2. 
\end{eqnarray}
\noindent
The couplings appearing in the chargino $\tilde{\chi}^+_j$ exchange 
are combined to:  
\begin{eqnarray}
   J'_j = - \frac{1}{4} \, (O^R_{0 j} - O^L_{0 j}) \,
   (Q'^R_{0 j} \, \sin \beta + Q'^L_{0 j} \, \cos \beta) 
\end{eqnarray}
\begin{eqnarray}
   J''_j = - \frac{1}{4} \, (O^R_{0 j} + O^L_{0 j}) \,
   (Q'^R_{0 j} \, \sin \beta - Q'^L_{0 j} \, \cos \beta) 
\end{eqnarray}
\begin{eqnarray}
   K'_j = - \frac{1}{4} \, (O^R_{0 j} - O^L_{0 j}) \,
   (Q'^R_{0 j} \, \sin \beta - Q'^L_{0 j} \, \cos \beta) 
\end{eqnarray}
\begin{eqnarray}
   K''_j = - \frac{1}{4} \, (O^R_{0 j} + O^L_{0 j}) \,
   (Q'^R_{0 j} \, \sin \beta + Q'^L_{0 j} \, \cos \beta). 
\end{eqnarray}
\noindent
$L_i$ is the product of the coupling of the scalar Higgs $S_i$
to the LSP neutralinos and to the $W$ boson and 
the charged Higgs.
In the NMSSM $L_i$ is given by: 
\begin{eqnarray}
   L_i = \frac{1}{2} \, 
   (U^S_{i 2} \, \cos \beta - U^S_{i 1} \, \sin \beta) \, 
   T_{S_i 0 0}.  
\end{eqnarray}
\noindent
In the NMSSM the coupling $g_{\alpha - +}$ of the pseudoscalar Higgs 
$P_{\alpha}$ to the W boson and charged Higgs reads:
\begin{eqnarray}
   g_{\alpha - +} = U^P_{\alpha 2} \, \cos \beta 
   + U^P_{\alpha 1} \, \sin \beta.  \label{gamp}
\end{eqnarray}
\noindent
In the MSSM with only one pseudoscalar Higgs $P$ 
the coupling $g_{\alpha - +}$ 
turns out to \mbox{be 1 \cite{drees}}.
\newpage

\[ 6. \:\:\: \tilde{\chi}^0_1 \: \tilde{\chi}^0_1 \rightarrow 
   S_a \: S_b \:\: or \:\: P_{\alpha} \: P_{\beta}. \]
\noindent
The scalar Higgses $S_a$ and $S_b$ 
are produced by scalar Higgs $S_i \: (i = 1, 2, 3)$
exchange in the s-channel and neutralino 
$\tilde{\chi}^0_j \:  (j = 1-5)$ exchange in the t- and u-channel. 

\vspace*{.2cm}
\noindent
$ A(^3P_0): $
\begin{eqnarray}
   \sqrt{6} \, v \, g^2_2 \, \left[ g_{a b i} \, T_{S_i 0 0}
   \frac{R_Z}{4 - R_{S_i}^2 + i G_{S_i}}
   - 2 \, T_{S_a 0 j} \, T_{S_b 0 j} \frac{1 + R_j}{P_j}
   + \frac{4}{3} \, T_{S_a 0 j} \, T_{S_b 0 j} \,
   \frac{\overline{\beta}_f^2}{P_j^2} \right]  
\end{eqnarray}

\noindent
$ A(^3P_2): $
\begin{eqnarray}
   - \frac{8}{\sqrt{3}} \, v \, \overline{\beta}_f^2 \, g^2_2 \,
   T_{S_a 0 j} \, T_{S_b 0 j} \, \frac{1}{P_j^2} 
\end{eqnarray}

\noindent
The pseudoscalar Higgses $P_{\alpha}$ and $P_{\beta}$ are 
also produced by scalar Higgs $S_i \: (i = 1, 2, 3)$
exchange and neutralino $\tilde{\chi}^0_j \:  (j = 1-5)$ exchange.
The corresponding amplitudes for $P_{\alpha}$ and $P_{\beta}$
production can be obtained by replacing $a, b$ by $\alpha, \beta$
or $S_a, S_b$ by $P_{\alpha}, P_{\beta}$ and $R_j$ by $- R_j$.
In the NMSSM the trilinear scalar Higgs coupling 
$g_{a a a}$ is given by: 
\[ g_{a a a} = - \frac{6}{g_2 \, m_Z} \, \left(
   \frac{g^2_y + g^2_2}{8} \, \sqrt{2} \, \left(
   v_1 \, U^{S \, 3}_{a 1} + v_2 \, U^{S \, 3}_{a 2} 
   - v_1 \, U^S_{a 1} \, U^{S \, 2}_{a 2} 
   - v_2 \, U^S_{a 2} \, U^{S \, 2}_{a 1} \right) \right. \]
\[ + \frac{\lambda^2}{\sqrt{2}} \, v_2 \, U^S_{a 2} \, U^{S \, 2}_{a 3}  
   + \frac{\lambda^2}{\sqrt{2}} \, x \, U^S_{a 3} \, U^{S \, 2}_{a 2}   
   + \frac{\lambda^2}{\sqrt{2}} \, v_1 \, U^S_{a 1} \, U^{S \, 2}_{a 3}  
   + \frac{\lambda^2}{\sqrt{2}} \, x \, U^S_{a 3} \, U^{S \, 2}_{a 1} \]  
\[ + \frac{\lambda^2}{\sqrt{2}} \, v_1 \, U^S_{a 1} \, U^{S \, 2}_{a 2}  
   + \frac{\lambda^2}{\sqrt{2}} \, v_2 \, U^S_{a 2} \, U^{S \, 2}_{a 1}   
   + \sqrt{2} \, k^2 \, x \, U^{S \, 3}_{a 3} \] 
\[ - \frac{\lambda \, k}{\sqrt{2}} \, v_1 \, U^S_{a 2} \, U^{S \, 2}_{a 3} 
   - \frac{\lambda \, k}{\sqrt{2}} \, 
   v_2 \, U^S_{a 1} \, U^{S \, 2}_{a 3}   
   - \sqrt{2} \, \lambda \, k \, x \, U^S_{a 1} \, U^S_{a 2} \, U^S_{a 3} \]
\begin{eqnarray}
   \left. - \frac{\lambda \, A_{\lambda}}{\sqrt{2}} \, 
   U^S_{a 1} \, U^S_{a 2} \, U^S_{a 3}   
   - \frac{k \, A_k}{3 \, \sqrt{2}} \, 
   U^{S \, 3}_{a 3} \right).   
\end{eqnarray}
\noindent
The trilinear scalar Higgs coupling $g_{a a b}$ (or $g_{a b a}$) 
has the following form: 
\[ g_{a a b} = g_{a b a} = - \frac{2}{g_2 \, m_Z} \, \left(
   \frac{g^2_y + g^2_2}{8} \, \sqrt{2} \, \left(
   3 \, v_1 \, U^S_{b 1} \, U^{S \, 2}_{a 1} 
   + 3 \, v_2 \, U^S_{b 2} \, U^{S \, 2}_{a 2} 
   - v_1 \, U^S_{b 1} \, U^{S \, 2}_{a 2} \right. \right. \]
\[ \left. - 2 \, v_1 \, U^S_{a 1} \, U^S_{b 2} \, U^S_{a 2} 
   - v_2 \, U^S_{b 2} \, U^{S \, 2}_{a 1}
   - 2 \, v_2 \, U^S_{a 2} \, U^S_{b 1} \, U^S_{a 1} \right) \]
\[ + \frac{\lambda^2 \, v_2 - \lambda \, k \, v_1}{\sqrt{2}} \,
   \left( U^S_{b 2} \, U^{S \, 2}_{a 3} 
   + 2 \, U^S_{a 2} \, U^S_{b 3} \, U^S_{a 3} \right) 
   + \frac{\lambda^2 \, x}{\sqrt{2}} \,
   \left( U^S_{b 3} \, U^{S \, 2}_{a 2} 
   + 2 \, U^S_{a 3} \, U^S_{b 2} \, U^S_{a 2} \right) \]
\[ + \frac{\lambda^2 \, v_1 - \lambda \, k \, v_2}{\sqrt{2}} \,
   \left( U^S_{b 1} \, U^{S \, 2}_{a 3} 
   + 2 \, U^S_{a 1} \, U^S_{b 3} \, U^S_{a 3} \right) 
   + \frac{\lambda^2 \, x}{\sqrt{2}} \,
   \left( U^S_{b 3} \, U^{S \, 2}_{a 1} 
   + 2 \, U^S_{a 3} \, U^S_{b 1} \, U^S_{a 1} \right) \]
\[ + \frac{\lambda^2 \, v_1}{\sqrt{2}} \,
   \left( U^S_{b 1} \, U^{S \, 2}_{a 2} 
   + 2 \, U^S_{a 1} \, U^S_{b 2} \, U^S_{a 2} \right)
   + \frac{\lambda^2 \, v_2}{\sqrt{2}} \,
   \left( U^S_{b 2} \, U^{S \, 2}_{a 1} 
   + 2 \, U^S_{a 2} \, U^S_{b 1} \, U^S_{a 1} \right) \]
\[ + \frac{6 \, k^2 \, x - k \, A_k}{\sqrt{2}} \, 
   U^S_{b 3} \, U^{S \, 2}_{a 3} \]
\begin{eqnarray}
   \left.  
   - \frac{2 \, \lambda \, k \, x + \lambda \, A_{\lambda}}{\sqrt{2}} \, 
   \left( U^S_{b 1} \, U^S_{a 2} \, U^S_{a 3}
   + U^S_{a 1} \, U^S_{b 2} \, U^S_{a 3}
   + U^S_{a 1} \, U^S_{a 2} \, U^S_{b 3} \right) \right). 
\end{eqnarray}
\noindent
Finally the trilinear scalar Higgs coupling $g_{1 2 3}$ 
(or $g_{1 3 2}$ or $g_{2 3 1}$) reads: 
\[ g_{1 2 3} = g_{1 3 2} = g_{2 3 1} = - \frac{1}{g_2 \, m_Z} \, \left(
   \frac{g^2_y + g^2_2}{8} \, \sqrt{2} \, \left(
   6 \, v_1 \, U^S_{1 1} \, U^S_{2 1} \, U^S_{3 1}
   + 6 \, v_2 \, U^S_{1 2} \, U^S_{2 2} \, U^S_{3 2} \right. \right.\]
\[ - 2 \, v_1 \, \left( U^S_{1 1} \, U^S_{2 2} \, U^S_{3 2} 
   + U^S_{2 1} \, U^S_{1 2} \, U^S_{3 2} 
   + U^S_{3 1} \, U^S_{1 2} \, U^S_{2 2} \right) \]
\[ \left. - 2 \, v_2 \, \left( U^S_{1 2} \, U^S_{2 1} \, U^S_{3 1} 
   + U^S_{2 2} \, U^S_{1 1} \, U^S_{3 1} 
   + U^S_{3 2} \, U^S_{1 1} \, U^S_{2 1} \right) \right) \]
\[ + \sqrt{2}\, \left( \lambda^2 \, v_2 - \lambda \, k \, v_1 \right) \,
   \left( U^S_{1 2} \, U^S_{2 3} \, U^S_{3 3}
   + U^S_{2 2} \, U^S_{1 3} \, U^S_{3 3}
   + U^S_{3 2} \, U^S_{1 3} \, U^S_{2 3} \right) \]
\[ + \sqrt{2}\, \lambda^2 \, x \,
   \left( U^S_{1 3} \, U^S_{2 2} \, U^S_{3 2}
   + U^S_{2 3} \, U^S_{1 2} \, U^S_{3 2}
   + U^S_{3 3} \, U^S_{1 2} \, U^S_{2 2} \right) \]
\[ + \sqrt{2}\, \left( \lambda^2 \, v_1 - \lambda \, k \, v_2 \right) \,
   \left( U^S_{1 1} \, U^S_{2 3} \, U^S_{3 3}
   + U^S_{2 1} \, U^S_{1 3} \, U^S_{3 3}
   + U^S_{3 1} \, U^S_{1 3} \, U^S_{2 3} \right) \]
\[ + \sqrt{2}\, \lambda^2 \, x \,
   \left( U^S_{1 3} \, U^S_{2 1} \, U^S_{3 1}
   + U^S_{2 3} \, U^S_{1 1} \, U^S_{3 1}
   + U^S_{3 3} \, U^S_{1 1} \, U^S_{2 1} \right) \]
\[ + \sqrt{2}\, \lambda^2 \, v_1 \,
   \left( U^S_{1 1} \, U^S_{2 2} \, U^S_{3 2}
   + U^S_{2 1} \, U^S_{1 2} \, U^S_{3 2}
   + U^S_{3 1} \, U^S_{1 2} \, U^S_{2 2} \right) \]
\[ + \sqrt{2}\, \lambda^2 \, v_2 \,
   \left( U^S_{1 2} \, U^S_{2 1} \, U^S_{3 1}
   + U^S_{2 2} \, U^S_{1 1} \, U^S_{3 1}
   + U^S_{3 2} \, U^S_{1 1} \, U^S_{2 1} \right) \]
\[ + \sqrt{2}\, \left( 6 \, k^2 \, x - k \, A_k \right) \,
   U^S_{1 3} \, U^S_{2 3} \, U^S_{3 3} \]
\[  - \frac{2 \, \lambda \, k \, x + \lambda \, A_{\lambda}}
   {\sqrt{2}} \, \left( U^S_{1 1} \, U^S_{2 2} \, U^S_{3 3}
   + U^S_{1 1} \, U^S_{3 2} \, U^S_{2 3}
   + U^S_{2 1} \, U^S_{1 2} \, U^S_{3 3} \right. \]
\begin{eqnarray}
   \left. + U^S_{2 1} \, U^S_{3 2} \, U^S_{1 3}
   + U^S_{3 1} \, U^S_{1 2} \, U^S_{2 3}
   + U^S_{3 1} \, U^S_{2 2} \, U^S_{1 3} \right) \Bigg). 
\end{eqnarray}
\noindent
The trilinear scalar and pseudoscalar Higgs coupling 
$g_{\alpha \alpha a}$ (or $g_{a \alpha \alpha}$) is given by:
\[ g_{\alpha \alpha a} = g_{a \alpha \alpha} = 
   - \frac{2}{g_2 \, m_Z} \, \left(
   \frac{g^2_y + g^2_2}{8} \, \sqrt{2} \, \left(
   v_1 \, U^S_{a 1} \, U^{P \, 2}_{\alpha 1}
   + v_2 \, U^S_{a 2} \, U^{P \, 2}_{\alpha 2} \right. \right. \]
\[ \left. - v_1 \, U^S_{a 1} \, U^{P \, 2}_{\alpha 2}
   - v_2 \, U^S_{a 2} \, U^{P \, 2}_{\alpha 1} \right) 
   + \frac{\lambda^2}{\sqrt{2}} \, \left(
   v_2 \, U^S_{a 2} \, U^{P \, 2}_{\alpha 3}
   + x \, U^S_{a 3} \, U^{P \, 2}_{\alpha 2}
   + v_1 \, U^S_{a 1} \, U^{P \, 2}_{\alpha 3} \right. \]
\[ \left. + x \, U^S_{a 3} \, U^{P \, 2}_{\alpha 1} 
   + v_1 \, U^S_{a 1} \, U^{P \, 2}_{\alpha 2}
   + v_2 \, U^S_{a 2} \, U^{P \, 2}_{\alpha 1} \right) \]
\[ + \sqrt{2} \, k^2 \, x \, U^S_{a 3} \, U^{P \, 2}_{\alpha 3} 
   + \frac{\lambda \, k}{\sqrt{2}} \, \left( 
   v_1 \, U^S_{a 2} \, U^{P \, 2}_{\alpha 3}
   + v_2 \, U^S_{a 1} \, U^{P \, 2}_{\alpha 3} \right) \]
\[ - \sqrt{2} \, \lambda \, k \, \left(
   v_1 \, U^S_{a 3} \, U^P_{\alpha 3} \, U^P_{\alpha 2}
   + v_2 \, U^S_{a 3} \, U^P_{\alpha 3} \, U^P_{\alpha 1} \right. \]  
\[ \left. + x \, U^S_{a 1} \, U^P_{\alpha 2} \, U^P_{\alpha 3} 
   + x \, U^S_{a 2} \, U^P_{\alpha 1} \, U^P_{\alpha 3} 
   - x \, U^S_{a 3} \, U^P_{\alpha 1} \, U^P_{\alpha 2} \right) \] 
\begin{eqnarray}
   + \frac{\lambda \, A_{\lambda}}{\sqrt{2}} \, \left(
   U^S_{a 1} \, U^P_{\alpha 2} \, U^P_{\alpha 3}
   + U^S_{a 2} \, U^P_{\alpha 1} \, U^P_{\alpha 3}   
   + U^S_{a 3} \, U^P_{\alpha 1} \, U^P_{\alpha 2} \right) 
   + \frac{k \, A_k}{\sqrt{2}} \, 
   U^S_{a 3} \, U^{P 2}_{\alpha 3} \Bigg). 
\end{eqnarray}
\noindent
The trilinear scalar and pseudoscalar Higgs coupling 
$g_{\alpha \beta a}$ (or $g_{a \alpha \beta}$ or $g_{a \beta \alpha}$) 
can be written as:
\[ g_{\alpha \beta a} = g_{a \alpha \beta} = g_{a \beta \alpha} = 
   - \frac{1}{g_2 \, m_Z} \, \left(
   \frac{g^2_y + g^2_2}{8} \, \sqrt{2} \, \left(
   v_1 \, U^S_{a 1} \, U^P_{\alpha 1} \, U^P_{\beta 1} \right. \right. \]
\[ \left. + v_2 \, U^S_{a 2} \, U^P_{\alpha 2} \, U^P_{\beta 2}
   - v_1 \, U^S_{a 1} \, U^P_{\alpha 2} \, U^P_{\beta 2}
   - v_2 \, U^S_{a 2} \, U^P_{\alpha 1} \, U^P_{\beta 1} \right) \] 
\[ + \sqrt{2} \, \lambda^2 \, \left(
   v_2 \, U^S_{a 2} \, U^P_{\alpha 3} \, U^P_{\beta 3} 
   + x \, U^S_{a 3} \, U^P_{\alpha 2} \, U^P_{\beta 2} 
   + v_1 \, U^S_{a 1} \, U^P_{\alpha 3} \, U^P_{\beta 3} 
\right. \]
\[ \left. + x \, U^S_{a 3} \, U^P_{\alpha 1} \, U^P_{\beta 1} 
   + v_1 \, U^S_{a 1} \, U^P_{\alpha 2} \, U^P_{\beta 2} 
   + v_2 \, U^S_{a 2} \, U^P_{\alpha 1} \, U^P_{\beta 1} \right) 
   + 2 \, \sqrt{2} \, k^2 \, x \, 
   U^S_{a 3} \, U^P_{\alpha 3} \, U^P_{\beta 3} \]
\[ + \sqrt{2} \, \lambda k \, v_1 \, 
   U^S_{a 2} \, U^P_{\alpha 3} \, U^P_{\beta 3} 
   - \sqrt{2} \, \lambda k \, v_1 \, U^S_{a 3} \, 
   \left( U^P_{\alpha 3} \, U^P_{\beta 2} 
   + U^P_{\beta 3} \, U^P_{\alpha 2} \right) \]
\[ + \sqrt{2} \, \lambda k \, v_2 \, 
   U^S_{a 1} \, U^P_{\alpha 3} \, U^P_{\beta 3} 
   - \sqrt{2} \, \lambda k \, v_2 \, U^S_{a 3} \, 
   \left( U^P_{\alpha 3} \, U^P_{\beta 1} 
   + U^P_{\beta 3} \, U^P_{\alpha 1} \right) \]
\[ + \frac{- 2 \, \lambda \, k \, x + \lambda \, A_{\lambda}}
   {\sqrt{2}} \, \bigg( U^S_{a 1} \, 
   \left( U^P_{\alpha 2} \, U^P_{\beta 3} 
   + U^P_{\beta 2} \, U^P_{\alpha 3} \right) 
   + U^S_{a 2} \, \left( U^P_{\alpha 1} \, U^P_{\beta 3} 
   + U^P_{\beta 1} \, U^P_{\alpha 3} \right) \bigg) \]
\begin{eqnarray}
   + \frac{2 \, \lambda \, k \, x + \lambda \, A_{\lambda}}
   {\sqrt{2}} \, U^S_{a 3} \, 
   \left( U^P_{\alpha 1} \, U^P_{\beta 2} 
   + U^P_{\beta 1} \, U^P_{\alpha 2} \right) 
   + \sqrt{2} \, k \, A_k \, 
   U^S_{a 3} \, U^P_{\alpha 3} \, U^P_{\beta 3} \Bigg).  
\end{eqnarray}

\[ 7. \:\:\: \tilde{\chi}^0_1 \: \tilde{\chi}^0_1 \rightarrow 
   S_a \: P_{\beta}. \]
\noindent
The scalar Higgs $S_a$ and the pseudoscalar Higgs $P_{\beta}$
are produced by Z boson and pseudoscalar Higgs 
$P_{\alpha} \: (\alpha = 1, 2)$ exchange in the s-channel
and neutralino $\tilde{\chi}^0_j \:  (j = 1-5)$ exchange 
in the t- and u-channel. 

\vspace*{.2cm}
\noindent
$ A(^1S_0): $
\[ 2 \, \sqrt{2} \, g^2_2 \, g_{a \alpha \beta} T_{P_{\alpha} 0 0}
   \frac{R_Z}{4 - R_{P_{\alpha}}^2}
   \left[ 1 + \frac{v^2}{8} \right]
   + \sqrt{2} \, g^2_2 \, g_{a \beta Z} 
   \frac{R_{P_{\beta}}^2 - R_{S_{a}}^2}{R_Z^2}
   \left[ 1 - \frac{v^2}{8} \right] \]
\[ - 4 \, \sqrt{2} \, g^2_2 \, T_{S_a 0 j} \, T_{P_{\beta} 0 j} 
   \frac{R_j}{P_j} \left[ 1 + v^2 \left[ \frac{1}{8} 
   - \frac{1}{2 P_j} + \frac{\overline{\beta}_f^2}{3 P_j^2} 
   \right] \right] \] 
\begin{eqnarray}
   - \sqrt{2} \, (R_{P_{\beta}}^2 - R_{S_{a}}^2) \, g^2_2 \, 
   T_{S_a 0 j} \, T_{P_{\beta} 0 j} \left[ 1 + v^2 \left[ - \frac{1}{8} 
   - \frac{1}{2 P_j} + \frac{\overline{\beta}_f^2}{3 P_j^2} 
   \right] \right] 
\end{eqnarray}

$ A(^3P_1): $
\begin{eqnarray}
   - 4 \, v \, \overline{\beta}_f
   \frac{g^2_2 \, g_{a \beta Z}}{4 - R_Z^2 + i G_Z}
   - 4 \, v \, \overline{\beta}_f g^2_2
   \frac{T_{S_a 0 j} \, T_{P_{\beta} 0 j}}{P_j}  
\end{eqnarray}

\noindent
$G_Z \: = \: \Gamma_Z \: m_Z \: / \: m_{\tilde{\chi}}^2$ is 
the rescaled widths of the Z boson.
The couplings appearing in the Z boson exchange are combined to 
$g_{a \beta Z}$. 
In the NMSSM $g_{a \beta Z}$ has the following form: 
\begin{eqnarray}
   g_{a \beta Z} = - \frac{1}{2 \, \cos^2 \theta_W} \,
   (U^S_{a 2} \, U^P_{\beta 2} - U^S_{a 1} \, U^P_{\beta 1}) \, 
   O''^L_{0 0}. 
\end{eqnarray}

\[ 8. \:\:\: \tilde{\chi}^0_1 \: \tilde{\chi}^0_1 \rightarrow 
   H^+ \: H^-. \]
\noindent
The charged Higgses are produced by Z boson and scalar Higgs 
$S_i \: (i = 1, 2, 3)$ exchange in the s-channel 
and chargino $\tilde{\chi}^+_j \: (j = 1,2)$
exchange in the t- and u-channel.

\vspace*{.2cm}
\noindent
$ A(^3P_0): $
\[ - \sqrt{6} \, v \, g^2_2 \, \left[ (Q'^{L 2}_{0 j} + Q'^{R 2}_{0 j})
   \left[ \frac{1}{P_j} - \frac{2 \, \overline{\beta}_f^2}{3 P_j^2} \right] 
   + 2 \, Q'^L_{0 j} \, Q'^R_{0 j} \frac{R^+_j}{P_j} \right] \]
\begin{eqnarray}
   + \sqrt{6} \, v \, g^2_2 \, R_W \, \frac{g_{+ - i}}{4 - R^2_{S_i}} 
\end{eqnarray}

\noindent
$ A(^3P_1): $
\begin{eqnarray}
   2 \, v \, \overline{\beta}_f \, g^2_2 \, 
   (Q'^{L 2}_{0 j} - Q'^{R 2}_{0 j}) \, \frac{1}{P_j}
   + 4 \, v \, \overline{\beta}_f \, 
   \frac{\cos (2 \theta_W)}{\cos^2 \theta_W} \, g^2_2
   \frac{O''^L_{00}}{4 - R_Z^2} 
\end{eqnarray}

\noindent
$ A(^3P_2): $
\begin{eqnarray}
   \frac{4}{\sqrt{3}} \, v \, \overline{\beta}_f^2 \, g^2_2 \, 
   (Q'^{L 2}_{0 j} + Q'^{R 2}_{0 j}) \, \frac{1}{P^2_j} 
\end{eqnarray}

\noindent
$g_{+ - i}$ is the product of the coupling of the scalar Higgs $S_i$ 
to the LSP neutralinos and to the charged Higgses.
In the NMSSM $g_{+ - i}$ is given by:
\[ g_{+ - i} = - \, T_{S_i 0 0} \, \Bigg[
   (U^S_{i 2} \, \sin \beta + U^S_{i 1} \, \cos \beta) 
   + \frac{\cos (2 \beta)}{2 \, \cos^2 \theta_W} \,
   (- U^S_{i 1} \, \cos \beta + U^S_{i 2} \, \sin \beta) \]
\[ \mbox{} - \frac{\sqrt{2} \, \lambda^2 v}{g_2 \, m_W} \,
   (U^S_{i 2} \, \sin \beta \, \cos^2 \beta 
   + U^S_{i 1} \, \sin^2 \beta \, \cos \beta) 
   + \frac{\sqrt{2} \, \lambda^2 x}{g_2 \, m_W} \, U^S_{i 3} \]
\begin{eqnarray}
   \mbox{} + \frac{2 \, \sqrt{2} \, \lambda \, k \, x}{g_2 \, m_W} \, 
   U^S_{i 3} \, \sin \beta \, \cos \beta 
   + \frac{\sqrt{2} \, \lambda \, A_{\lambda}}{g_2 \, m_W} \, 
   U^S_{i 3} \, \sin \beta \, \cos \beta \Bigg]. 
\end{eqnarray}

\[ 9. \:\:\: \tilde{\chi}^0_1 \: \tilde{\chi}^0_1 \rightarrow 
   f_a(h) \: \overline{f}_a(\overline{h}). \]
\noindent
The fermions are produced by Z boson, scalar Higgs $S_i \: (i = 1, 2, 3)$
and pseudoscalar Higgs $P_{\alpha} \: (\alpha = 1, 2)$ exchange in the 
s-channel and sfermion $\tilde{f}_{1, 2}$ exchange in the t- and u-channel.

\vspace*{.2cm}
\noindent
$ A(^1S_0): \lambda_f = 0 $
\[ \sqrt{2} \, (-1)^{\overline{h}+1/2} \, ({X'}_{a 0}^2 + {W'}_{a 0}^2) 
   \left[ 1 + v^2 \left[ - \frac{1}{2 \, P_1} + 
     \frac{\overline{\beta}_f^2}{3 \, P_1^2} \right] \right] 
   \frac{R_f}{P_1} \]
\[ + 2 \, \sqrt{2} \, (-1)^{\overline{h}+1/2} \, X'_{a 0} \, W'_{a 0} \,
   \frac{1}{P_1} \left[ 1 + v^2 \left[ \frac{1}{4} - \frac{1}{2 \, P_1} - 
   \frac{\overline{\beta}_f^2}{6 \, P_1} +
   \frac{\overline{\beta}_f^2}{3 \, P_1^2} \right] \right] \]
\[ + (X'_{a0} \leftrightarrow Z'_{a0},
      W'_{a0} \leftrightarrow Y'_{a0}, P_1 \leftrightarrow P_2) 
   + (-1)^{\overline{h}+1/2} \, \frac{2 \, \sqrt{2} \, g^2_2}
   {\cos^2 \theta_W} \, {O''}_{00}^L \, T_{3 a} \, \frac{R_f}{R_Z^2} \]
\begin{eqnarray}
   + 4 \, \sqrt{2} \, (-1)^{\overline{h}+1/2} \, g_2 \, 
   h_{P_{\alpha} a} \, T_{P_{\alpha} 0 0} \, 
   \frac{1}{4 - R_{P_{\alpha}}^2 + i G_{P_{\alpha}}} 
   \left[1 + \frac{v^2}{4}\right] 
\end{eqnarray}
  
\noindent
$ A(^3P_0): \lambda_f = 0 $
\[ - \sqrt{6} \, v \, \overline{\beta}_f \, (X'_{a0} W'_{a0}) 
   \left[ \frac{1}{P_1} - \frac{2}{3 P_1^2} \right]
   + \frac{\sqrt{6}}{3} \, v \, \overline{\beta}_f \, 
   ({X'}_{a0}^2 + {W'}_{a0}^2) \frac{R_f}{P_1^2} \]
\begin{eqnarray}
   + (X'_{a0} \leftrightarrow Z'_{a0},
      W'_{a0} \leftrightarrow Y'_{a0}, P_1 \leftrightarrow P_2)  
   - 2 \, \sqrt{6} \, v \, \overline{\beta}_f \, g_2 \,
   \frac{h_{S_i a} \, T_{S_i 0 0}}{4 - R_{S_i}^2 + i G_{S_i}} 
\end{eqnarray}

\noindent
$ A(^3P_1): \lambda_f = 0 $
\[ \frac{v \, R_f}{P_1} ({X'}_{a0}^2 - {W'}_{a0}^2)
   + (X'_{a0} \leftrightarrow Z'_{a0},
      W'_{a0} \leftrightarrow Y'_{a0}, P_1 \leftrightarrow P_2) \]
\begin{eqnarray}
   - 2 \, v \, g^2_2 \, \frac{{O''}_{00}^L}{\cos^2 \theta_W}
   \left[ T_{3 a} - 2 \, e_{f_a} \, \sin^2 \theta_W \right]
   \frac{R_f}{4 - R_Z^2 + i G_Z} 
\end{eqnarray}
$ \lambda_f = \pm 1 $
\[ \sqrt{2} \, v \, \Bigg[ \lambda_f \, \overline{\beta}_f \,
   ({X'}_{a0}^2 + {W'}_{a0}^2) 
   \left[ - \frac{1}{P_1} + \frac{1}{P_1^2} \right]
   + ({X'}_{a0}^2 - {W'}_{a0}^2)
   \left[ \frac{1}{P_1} - \frac{\overline{\beta}_f^2}{P_1^2} \right] \]
\[ + 2 \, \overline{\beta}_f \, \lambda_f \, X'_{a0} \, W'_{a0} \,
   \frac{R_f}{P_1^2} \Bigg] 
   + (X'_{a0} \leftrightarrow Z'_{a0},
   W'_{a0} \leftrightarrow Y'_{a0}, P_1 \leftrightarrow P_2) \]
\begin{eqnarray}
   + 2 \, \sqrt{2} \, v \, \frac{g^2_2}{\cos^2 \theta_W} \,
   {O''}_{00}^L \, \left[ \lambda_f \, T_{3 a} \, \overline{\beta}_f
   - (T_{3 a} - 2 \, e_{f_a} \, \sin^2 \theta_W) \right]
   \frac{1}{4 - R_Z^2 + i G_Z} 
\end{eqnarray}

\noindent
$ A(^3P_2): \lambda_f = 0 $
\[ - \frac{2}{\sqrt{3}} \, v \, \overline{\beta}_f \, \frac{1}{P_1^2} \,
   \left[ R_f \, ({X'}_{a0}^2 + {W'}_{a0}^2) 
   + 2 \, X'_{a0} \, W'_{a0} \right] \]
\begin{eqnarray}
   + (X'_{a0} \leftrightarrow Z'_{a0},
   W'_{a0} \leftrightarrow Y'_{a0}, P_1 \leftrightarrow P_2) 
\end{eqnarray}
$ \lambda_f = \pm 1 $
\[ \sqrt{2} \, v \, \overline{\beta}_f \, \frac{1}{P_1^2} \,
   [ - ({X'}_{a0}^2 + {W'}_{a0}^2) 
   + \overline{\beta}_f \, \lambda_f \, ({X'}_{a0}^2 - {W'}_{a0}^2)
   - 2 \, R_f \, X'_{a0} \,W'_{a0} ] \]
\begin{eqnarray}
   + (X'_{a0} \leftrightarrow Z'_{a0},
   W'_{a0} \leftrightarrow Y'_{a0}, P_1 \leftrightarrow P_2). 
\end{eqnarray}

\noindent
The sfermion mass eigenstates $\tilde{f}_{1, 2}$
are defined by
\begin{eqnarray}
   \tilde{f}_1 = \tilde{f}_L \, \cos \theta_f 
   + \tilde{f}_R \, \sin \theta_f 
\end{eqnarray}
\begin{eqnarray}
   \tilde{f}_2 = - \tilde{f}_L \, \sin \theta_f 
   + \tilde{f}_R \, \cos \theta_f. 
\end{eqnarray}
\noindent
The couplings of the mixed sfermion $\tilde{f}_{1, 2}$ to the fermion 
and LSP neutralino can be written as: 
\begin{eqnarray}
   X'_{a 0} = X_{a 0} \, \cos \theta_f  
   + Z_{a 0} \, \sin \theta_f 
\end{eqnarray}
\begin{eqnarray}
   W'_{a 0} = Z_{a 0} \, \cos \theta_f  
   + Y_{a 0} \, \sin \theta_f 
\end{eqnarray}
\begin{eqnarray}
   Z'_{a 0} = - X_{a 0} \, \sin \theta_f  
   + Z_{a 0} \, \cos \theta_f 
\end{eqnarray}
\begin{eqnarray}
   Y'_{a 0} = - Z_{a 0} \, \sin \theta_f  
   + Y_{a 0} \, \cos \theta_f, 
\end{eqnarray}
\noindent
where the couplings of the unmixed sfermion $\tilde{f}_{L, R}$ 
to the fermion and LSP neutralino are given by: 
\begin{eqnarray}
   X_{a 0} = - \sqrt{2} \, g_2 \, [T_{3 a} \, N_{0 2}
   - \tan \theta_W \, (T_{3 a} - e_{f_a}) \, N_{0 1}] 
\end{eqnarray}
\begin{eqnarray}
   Y_{a 0} = \sqrt{2} \, g_2 \, \tan \theta_W \, e_{f_a} \, N_{0 1} 
\end{eqnarray}
\begin{eqnarray}
   Z_{u 0} = - \frac{g_2 \, m_u}{\sqrt{2} \, \sin \beta \, m_W } \, 
   N_{0 4} 
\end{eqnarray}
\begin{eqnarray}
   Z_{d 0} = - \frac{g_2 \, m_d}{\sqrt{2} \, \cos \beta \, m_W } \, 
   N_{0 3}. 
\end{eqnarray}
\noindent
In the NMSSM the couplings of the Higgses to the fermions 
have to be modified compared to the MSSM:
\begin{eqnarray}
  h_{S_i u} = - \frac{g_2 \: m_u \: U_{i 2}^S}{2 \: \sin \beta \: m_W}  
  \hspace*{2.cm}
  h_{S_i d} = - \frac{g_2 \: m_d \: U_{i 1}^S}{2 \: \cos \beta \: m_W} 
\end{eqnarray}
\begin{eqnarray}
  h_{P_{\alpha} u} = 
  - \frac{g_2 \: m_u \: U_{{\alpha} 2}^P}{2 \: \sin \beta \: m_W} 
  \hspace*{2.cm}
  h_{P_{\alpha} d} = 
  - \frac{g_2 \: m_d \: U_{{\alpha} 1}^P}{2 \: \cos \beta \: m_W}. 
\end{eqnarray}
\newpage
\section*{Appendix A2: Renormalisation Group Equations}
\noindent
For completeness we also give here the 1 loop SUSY RGEs 
\cite{savoy} for the parameters in the NMSSM with our conventions.
$Q$ is the running mass scale in the variable 
$t \: = \: \log ( Q/M_{weak} )$. 
\begin{eqnarray}
\frac{d}{d \: t} \: g_y & = & \frac{11}{16 \: \pi^2} \: g_y^3  \\
\frac{d}{d \: t} \: g_2 & = & \frac{1}{16 \: \pi^2} \: g_2^3  \\
\frac{d}{d \: t} \: g_3 & = & - \: \frac{3}{16 \: \pi^2} \: g_3^3  \\
\frac{d}{d \: t} \: \lambda & = & \frac{1}{8 \: \pi^2} \: 
\left( \: k^2 \: + \: 2 \: \lambda^2 \: + \: \frac{3}{2} \: h_t^2 \:
- \: \frac{3}{2} \: g_2^2 \: - \: \frac{1}{2} \: g_y^2 \: \right) \: 
\lambda  \\
\frac{d}{d \: t} \: k & = & \frac{3}{8 \: \pi^2} \: 
\left( \: k^2 \: + \: \lambda^2 \: \right) \: k  \\
\frac{d}{d \: t} \: h_t & = & \frac{1}{8 \: \pi^2} \: 
\left( \: \frac{1}{2} \: \lambda^2 \: + \: 3 \: h_t^2 \:
- \: \frac{8}{3} \: g_3^2 \: - \: \frac{3}{2} \: g_2^2
- \: \frac{13}{18} \: g_y^2 \: \right) \: h_t  \\  
\frac{d}{d \: t} \: A_{\lambda} & = & \frac{1}{8 \: \pi^2} \: 
\left( \: 4 \: \lambda^2 \: A_{\lambda} \: 
- \: 2 \: k^2 \: A_k \: - \: 3 \: h_t^2 \: A_{U_3} \right. \nonumber  \\ 
& & \left. \mbox{}
- \: 3 \: g_2^2 \: M_2 \: - \: g_y^2 \: M_1 \: \right)  \\
\frac{d}{d \: t}  \: A_k & = & \frac{6}{8 \: \pi^2} \: 
\left( \: - \: \lambda^2 \: A_{\lambda} \: 
+ \: k^2 \: A_k \: \right) \\
\frac{d}{d \: t} \: A_{U_i} & = & \frac{1}{8 \: \pi^2} \: 
\Bigg( - \: \lambda^2 \: A_{\lambda} \: 
+ \: 3 \: ( \: 1 \: + \: \delta_{i 3} \: ) \: h_t^2 \: A_{U_3} 
\nonumber  \\ 
& & \left. \mbox{} 
+ \: \frac{16}{3} \: g_3^2 \: M_3 \: + \: 3 \: g_2^2 \: M_2 \: 
+ \: \frac{13}{9} \: g_y^2 \: M_1 \: \right)  \\
\frac{d}{d \: t} \: A_{D_i} & = & \frac{1}{8 \: \pi^2} \: 
\Bigg( - \: \lambda^2 \: A_{\lambda} \: 
+ \: \delta_{i 3} \: h_t^2 \: A_{U_3}  \nonumber  \\
& & \left. \mbox{} 
+ \: \frac{16}{3} \: g_3^2 \: M_3 \: + \: 3 \: g_2^2 \: M_2 \: 
+ \: \frac{7}{9} \: g_y^2 \: M_1 \: \right)  \\ 
\frac{d}{d \: t} \: A_{E_i} & = & \frac{1}{8 \: \pi^2} \: 
\left( - \: \lambda^2 \: A_{\lambda} \: 
+ \: 3 \: g_2^2 \: M_2 \: + \: 3 \: g_y^2 \: M_1 \: \right)  \\
\frac{d}{d \: t} \: m_N^2 & = & \frac{1}{8 \: \pi^2} \: 
\left[ 2 \: \lambda^2 \: \left( \: m_N^2 \: + \: m_{H_1}^2 \: + \: 
m_{H_2}^2 \: + \: A_{\lambda}^2 \: \right) \right. \nonumber  \\
& & \left. \mbox{} 
+ \: 2 \: k^2 \: \left( \: 3 \: m_N^2 \: + A_k^2 \: \right) \: \right]  \\ 
\frac{d}{d \: t} \: m_{H_1}^2 & = & \frac{1}{8 \: \pi^2} \: 
\left[ \lambda^2 \: \left( \: m_N^2 \: + \: m_{H_1}^2 \: + \: 
m_{H_2}^2 \: + \: A_{\lambda}^2 \: \right) \right. \nonumber  \\
& & \left. \mbox{} 
- \: 3 \: g_2^2 \: M_2^2 \: - \: g_y^2 \: M_1^2 \: \right]  \\
\frac{d}{d \: t} \: m_{H_2}^2 & = & \frac{d}{d \: t} \: m_{H_1}^2 \: 
+ \: \frac{3}{8 \: \pi^2} \: h_t^2 \: 
\left( \: m_{H_2}^2 \: + \: m_{U_3}^2 \: + \: m_{Q_3}^2 \: 
+ \: A_{U_3}^2 \: \right)  \\
\frac{d}{d \: t} \: m_{Q_i}^2 & = & \frac{1}{8 \: \pi^2} \: 
\Bigg[ \: \delta_{i 3} \: h_t^2 \: \left( \: m_{H_2}^2 \: 
+ \: m_{U_3}^2 \: + \: m_{Q_3}^2 \: + \: A_{U_3}^2 \: \right)   
\nonumber  \\
& & \left. \mbox{} 
- \: \frac{16}{3} \: g_3^2 \: M_3^2 \: 
- \: 3 \: g_2^2 \: M_2^2 \: 
- \: \frac{1}{9} \: g_y^2 \: M_1^2 \: \right]  \\
\frac{d}{d \: t} \: m_{U_i}^2 & = & \frac{1}{8 \: \pi^2} \: 
\Bigg[ \: 2 \: \delta_{i 3} \: h_t^2 \: \left( \: m_{H_2}^2 \: 
+ \: m_{U_3}^2 \: + \: m_{Q_3}^2 \: + \: A_{U_3}^2 \: \right)  
\nonumber  \\
& & \left. \mbox{}
- \: \frac{16}{3} \: g_3^2 \: M_3^2 \: 
- \: \frac{16}{9} \: g_y^2 \: M_1^2 \: \right]  \\ 
\frac{d}{d \: t} \: m_{D_i}^2 & = & \frac{1}{8 \: \pi^2} \: 
\left( \: - \: \frac{16}{3} \: g_3^2 \: M_3^2 \: 
- \: \frac{4}{9} \: g_y^2 \: M_1^2 \: \right)  \\ 
\frac{d}{d \: t} \: m_{L_i}^2 & = & \frac{1}{8 \: \pi^2} \: 
\left( \: - \: 3 \: g_2^2 \: M_2^2 \: 
- \: g_y^2 \: M_1^2 \: \right)  \\
\frac{d}{d \: t} \: m_{E_i}^2 & = & \frac{1}{8 \: \pi^2} \: 
\left( \: - \: 4 \: g_y^2 \: M_1^2 \: \right)  \\ 
\frac{d}{d \: t} \: M_1 & = & \frac{11}{8 \: \pi^2} \: 
g_y^2 \: M_1  \\
\frac{d}{d \: t} \: M_2 & = & \frac{1}{8 \: \pi^2} \: 
g_2^2 \: M_2  \\
\frac{d}{d \: t} \: M_3 & = & - \: \frac{3}{8 \: \pi^2} \: 
g_3^2 \: M_3 
\end{eqnarray}

\newpage

\section*{Figure Captions}
\begin{description}

\item[Fig\ 1] The scalar mass $m_0$ versus
the gaugino mass $|m_{1/2}|$. The upper pictures are for the MSSM,
the lower pictures are for the NMSSM. In the right pictures 
the dark matter condition is imposed.

\item[Fig\ 2] The trilinear coupling $|A_0|$ versus the 
scalar mass $m_0$. The upper pictures are for the MSSM,
the lower pictures are for the NMSSM. In the right pictures 
the dark matter condition is imposed.

\item[Fig\ 3] The gaugino mass $|M_2|$ versus the Higgs mixing
parameter $\mu$ both taken at the electroweak scale. 
The upper pictures are for the MSSM, the lower pictures 
are for the NMSSM. In the right pictures the dark matter condition 
is imposed.

\item[Fig\ 4] The bino portion of the LSP neutralino versus
the mass of the LSP neutralino. The upper pictures 
are for the MSSM, the lower pictures are for the NMSSM. 
In the right pictures the dark matter condition is imposed.

\item[Fig\ 5] The mass of the lighter chargino versus the 
lighter selectron mass. The upper pictures are for the MSSM, 
the lower pictures are for the NMSSM. 
In the right pictures the dark matter condition is imposed.

\item[Fig\ 6] The mass of the lighter u-squark versus the gluino mass. 
The upper pictures are for the MSSM, the lower pictures are for the NMSSM. 
In the right pictures the dark matter condition is imposed.

\item[Fig\ 7] The mass of the lighter stop versus the gluino mass. 
The upper pictures are for the MSSM, the lower pictures are for the NMSSM. 
In the right pictures the dark matter condition is imposed.

\item[Fig\ 8] The mass of the lightest scalar Higgs $S_1$ versus 
$|\tan \beta|$. The upper pictures are for the MSSM, the lower pictures are 
for the NMSSM. In the right pictures the dark matter condition is imposed.

\item[Fig\ 9] The mass of the lightest scalar Higgs $S_1$ versus 
the mass of the lightest pseudoscalar Higgs $P_1$.
The upper pictures are for the MSSM, the lower pictures are 
for the NMSSM. In the right pictures the dark matter condition is imposed.

\end{description}
\newpage

\newpage

\newpage

\newpage

\newpage

\newpage

\newpage

\newpage

\newpage

\newpage
\vspace*{.5cm}
\begin{thebibliography}{99}

\bibitem{nilles}
H.P. Nilles, Phys. Rep. 110 (1984) 1.

\bibitem{haber}
H.E. Haber, G.L. Kane, Phys. Rep. 117 (1985) 75;
\newline
J.F. Gunion, H.E. Haber, Nucl. Phys. B272 (1986) 1.

\bibitem{ellis}
J. Ellis, J.F. Gunion, H.E. Haber, L. Roszkowski, F. Zwirner,
\newline
Phys. Rev. D39 (1989) 844.

\bibitem{abelw}
S.A. Abel, S. Sarkar, P.L. White, Nucl. Phys. B454 (1995) 663;
\newline
S.A. Abel, Nucl. Phys. B480 (1996) 55.

\bibitem{ellwglmin}
U. Ellwanger, M. Rausch de Traubenberg, Z Phys. C53 (1992) 521.

\bibitem{ellwp}
U. Ellwanger, Phys. Lett. B303 (1993) 271.

\bibitem{ellwtnb}
U. Ellwanger, M. Rausch de Traubenberg, C.A. Savoy,
Phys. Lett. B315 (1993) 331.

\bibitem{ellwn}
U. Ellwanger, M. Rausch de Traubenberg, C.A. Savoy,
Z Phys. C67 (1995) 665.

\bibitem{ellws}
U. Ellwanger, M. Rausch de Traubenberg, C.A. Savoy,
Nucl. Phys. B492 (1997) 21.

\bibitem{ellwsi}
U. Ellwanger, C. Hugonie, Laboratoire de Physique Th\'{e}orique
et Hautes Energies Universit\'{e} de Paris-Sud,
Orsay preprint LPTHE-97-68, hep-ph/9712300.

\bibitem{elli1}
T. Elliott, S.F. King, P.L. White, Phys. Lett. B305 (1993) 71.

\bibitem{elli2}
T. Elliott, S.F. King, P.L. White, Phys. Lett. B314 (1993) 56;
\newline
Phys. Rev. D49 (1994) 2435.

\bibitem{king}
S.F. King, P.L. White, Phys. Rev. D52 (1995) 4183.

\bibitem{anan}
B. Ananthanarayan, P.N. Pandita, Phys. Lett. B371 (1996) 245.

\bibitem{franke}
F. Franke, H. Fraas, A. Bartl, Phys. Lett. B336 (1994) 415;
\newline
F. Franke, H. Fraas, Phys. Lett. B353 (1995) 234;
Z Phys. C72 (1996) 309.

\bibitem{greene}
B.R. Greene, P.J. Miron, Phys. Lett. B168 (1986) 226;
\newline
R. Flores, K.A. Olive, D. Thomas, Phys. Lett. B245 (1990) 509;
\newline
K.A. Olive, D. Thomas, Nucl. Phys. B355 (1991) 192.

\bibitem{abeln}
S.A. Abel, S. Sarkar, I.B. Whittingham, Nucl. Phys. B392 (1993) 83.

\bibitem{savoy}
J.P. Derendinger, C.A. Savoy, Nucl. Phys. B237 (1984) 307.

\bibitem{casas}
J.A. Casas, A. Lleyda, C. Mu\~{n}oz, Nucl. Phys. B471 (1996) 3;
\newline
J.A. Casas, Instituto de Estructura de la Materia, Madrid
\newline
preprint IEM-FT-161-97, hep-ph/9707475;
\newline
C. Mu\~{n}oz, Departamento de F\'{\i}sica Te\'{o}rica,
Universidad Aut\'{o}noma de Madrid,
\newline
preprint FTUAM-97-20, hep-ph/9709329.

\bibitem{ich}
A. Stephan, Phys. Lett. B411 (1997) 97.    

\bibitem{bartln}
A. Bartl, H. Fraas, W. Majerotto, Nucl. Phys. B278 (1986) 1.

\bibitem{ellisgut}
C. Giunti, C.W. Kim, U.W. Lee, Mod. Phys. Lett. A6 (1991) 1745;
\newline
J. Ellis, S. Kelley, D.V. Nanopoulos, Phys. Lett. B260 (1991) 131;
\newline
U. Amaldi, W. de Boer, H. F\"urstenau, Phys. Lett. B260 (1991) 447;
\newline
P. Langacker, M. Luo, Phys. Rev. D44 (1991) 817.

\bibitem{top}
M. Gallinaro, CDF Collab., preprint FERMILAB-CONF-97/004-E. 

\bibitem{PD}
Particle Data Group, R.M. Barnett et al., Phys. Rev. D54 (1996) 1. 

\bibitem{sfer}
DELPHI Collab., P. Andersson et al., Search for Sfermions at 
$\sqrt{s} = 161$ and $172 \, GeV$,
contribution to the HEP'97 Conference Jerusalem, August 19-26,
Ref.\#353.

\bibitem{bisset}
M. Bisset, S. Raychaudhuri, D.K. Ghosh,
Tata Institute of Fundamental Research,
Mumbai preprint TIFR-TH-96-45, hep-ph/9608421.

\bibitem{FERMILAB}
CDF Collab., F. Abe et al., Phys. Rev. D56 (1997) R1357.

\bibitem{charg}
L3 Collab., M. Acciarri et al., Preliminary results on 
New Particle Searches at $\sqrt{s} = 183 \, GeV$,
contribution to the HEP'97 Conference Jerusalem, August 19-26,
Ref.\#859.

\bibitem{chiggs}
OPAL Collab., K. Ackerstaff et al., Search for charged Higgs bosons
in $e^+ \: e^-$ collisions at   
$\sqrt{s} = 130 - 172 \, GeV$,
contribution to the HEP'97 Conference Jerusalem, August 19-26,
Ref.\#258.
  
\bibitem{higgs}
ALEPH Collab., R. Barate et al., Preliminary ALEPH results at 
$\sqrt{s} = 183 \, GeV$,
contribution to the HEP'97 Conference Jerusalem, August 19-26,
Ref.\#856.
  
\bibitem{L3}
L3 Collab., M. Acciarri et al., Phys. Lett. B385 (1996) 454;
\newline
ALEPH Collab., R. Barate et al., preprint CERN-PPE/97-071,
\newline
submitted to Phys. Lett. B.

\bibitem{diaz}
M.A. Diaz, S.F. King, Phys. Lett. B349 (1995) 105.

\bibitem{davis}
M. Davis, F. Summers, D. Schlegel, Nature 359 (1992) 393;
\newline
A. Klypin, J. Holtzman, J.R. Primack, E. Reg\"os, 
Astrophys. J. 416 (1993) 1.

\bibitem{drees}
M. Drees, M.M. Nojiri, Phys. Rev. D47 (1993) 376.

\bibitem{drees1}
M. Drees, Seoul National University,
preprint APCTP-96-04, hep-ph/9609300. 

\bibitem{kolb}
E.W. Kolb, M.S. Turner, The Early Universe 
(Addison-Wesley, New York, 1990).

\bibitem{griest}
K. Griest, M. Kamionkowski, M.S. Turner, Phys. Rev. D41 (1990) 3565.

\bibitem{griestv}
K. Griest, D. Seckel, Phys. Rev. D43 (1991) 3191;
\newline
P. Gondolo, G. Gelmini, Nucl. Phys. B360 (1991) 145;
\newline
P. Nath, R. Arnowitt, Phys. Rev. Lett. 70 (1993) 3696;
\newline
H. Baer, M. Brhlick, Phys. Rev. D53 (1996) 597.

\bibitem{edsjoe}
J. Edsj\"o, P. Gondolo, Phys. Rev. D56 (1997) 1879.

\bibitem{bartl}
A. Bartl, H. Fraas, W. Majerotto, B. M\"osslacher,
Z Phys. C55 (1992) 257.

\bibitem{kane}
G.L. Kane, C. Kolda, L. Roszkowski, J.D. Wells,  
Phys. Rev. D49 (1994) 6173.

\bibitem{kolda}
C. Kolda, L. Roszkowski, J.D. Wells, G.L. Kane, 
Phys. Rev. D50 (1994) 3498.

\bibitem{barger}
V. Barger, M.S. Berger, P. Ohmann, Phys. Rev. D49 (1994) 4908.

\bibitem{dreesp}
M. Drees, private communication.

\end{thebibliography}
\end{document}